\documentclass[structabstract]{aa}
\usepackage{subfigure}                                   % (traditional abstract) 
\usepackage{epsfig}
\usepackage{wasysym}
\usepackage{graphicx}
\usepackage{natbib}
\usepackage{lscape}
\usepackage{multirow}
\usepackage{xcolor}

\usepackage{booktabs}
  \def \teff {$T_{\mathrm{eff}}$}
  
  \def \vtur {$V_{\mathrm{tur}}$}
  \def \logg {$\log g$}

\begin{document}

  \title{The AMBRE project: searching for the closest solar siblings}

  \author{V.~Adibekyan\inst{1}
          \and P.~de~Laverny\inst{2}
          \and A.~Recio--Blanco\inst{2}
          \and S.~G.~Sousa\inst{1}
          \and E.~Delgado-Mena\inst{1}
          \and  \\G.~Kordopatis\inst{2}
          \and A.~C.~S.~Ferreira\inst{1}
          \and N.~C.~Santos\inst{1,3}
          \and A.~A.~Hakobyan\inst{4} 
          \and M.~Tsantaki\inst{1}
          }

  \institute{
          Instituto de Astrof\'isica e Ci\^encias do Espa\c{c}o, Universidade do Porto, CAUP, Rua das Estrelas, 4150-762 Porto, Portugal\\
          \email{vadibekyan@astro.up.pt}
          \and Université Côte d'Azur, Observatoire de la Côte d'Azur, CNRS, Laboratoire Lagrange, Bd de l'Observatoire, CS 34229, 06304 Nice cedex 4, France\
          \and Departamento de F\'isica e Astronomia, Faculdade de Ci\^encias, Universidade do Porto, Rua do Campo Alegre, 4169-007 Porto, Portugal\
          \and Byurakan Astrophysical Observatory, 0213 Byurakan, Aragatsotn province, Armenia
          }

  \date{Received date / Accepted date }
%----------------------------------------------------------------------------------------
%       Abstract
%----------------------------------------------------------------------------------------
  \abstract
  {Finding solar siblings, that is, stars that formed in the same cluster as the Sun, will yield information about the conditions at the Sun's birthplace.
  Finding possible solar siblings is difficult since they are spread widely throughout the Galaxy.}
  {We search for solar sibling candidates in AMBRE, the very large spectra database of solar vicinity stars.}
  {Since the ages and chemical abundances of solar siblings are very similar to those of the Sun, we carried out a chemistry- and age-based search for solar sibling candidates. We used high-resolution spectra to derive precise stellar parameters and chemical abundances of the stars. We used these spectroscopic parameters together with Gaia DR2 astrometric data to derive stellar isochronal ages. Gaia data were also used to study the kinematics of the sibling candidates.}
  {From the about 17\,000 stars that are characterized within the AMBRE project, we first selected 55 stars whose metallicities are closest to the solar value ($-0.1 \leq$ [Fe/H] $\leq$ 0.1 dex). For these stars we derived precise chemical abundances of several iron-peak, $\alpha$- and neutron-capture elements, based on  which we selected 12 solar sibling candidates with average abundances and metallicities between $-0.03$ to 0.03 dex. Our further selection left us with 4 candidates with stellar ages that are compatible with the solar age within observational uncertainties. For the 2 of the hottest candidates, we derived the carbon isotopic ratios, which are compatible with the solar value. HD186302 is the most precisely characterized and probably the most probable candidate of our 4 best candidates.}
  {Very precise chemical characterization and age estimation is necessary to identify solar siblings.  We propose that in addition to typical chemical tagging, the study of isotopic ratios can give further important information about the relation of sibling candidates with the Sun. Ideally, asteroseismic age determinations of the candidates could solve the problem of imprecise isochronal ages.}

%----------------------------------------------------------------------------------
%       Title
%----------------------------------------------------------------------------------

  \maketitle
%---------------------------------
   
%  -------------------------------------------------
%       Introduction
%----------------------------------------------------------------------------------
\section{Introduction}                                  \label{sec:Introduction}

As most low-mass stars, the Sun was probably formed in a cluster \citep[e.g.,][]{Lada-03} about 4.57 Gyr ago \citep{Bonanno-15}. The early solar nebula hosted short-lived radioactive isotopes (with half-lives shorter than a few Myr), which are mainly products of stellar nucleosynthesis. This indicates that  pollution by a supernova with a progenitor mass of $\sim$20 M$_{\odot}$ \citep[][]{Looney-06} or even higher \citep[e.g., $\sim$75-100 M$_\odot$][]{Williams-07} occurred in the first epochs ($\lesssim$ 2 Myr). The existence of such a massive star in the birth cluster of the Sun would require its initial mass to be at least 500 M$_\odot$ \citet{Weidner-04} or even 10$^{4}$ if the progenitor mass of the exploded supernova was $>$75 M$_\odot$ \citep{Weidner-04}.
  
Moreover, the dynamical excitement of the trans-Neptunian object Sedna  and the Kuiper belt objects \citep{Brown-04, Morbidelli-04} requires a close encounter with another star in the birth cluster, which also suggests that the solar birth cluster contained 10$^{3}$–10$^{4}$ stars \citep{Adams-10}. \citet{Zwart-09} suggested that the birth cluster could have had a mass between 500 and 3000 M$_\odot$ if the size of the cluster was 1--3 pc.

The lifetime of open clusters is typically about 200 Myr \citep[e.g.,][]{Piskunov-06}, although it depends on the cluster mass (massive clusters live longer) and on the galaxy properties \citep[e.g.,][]{Lamers-06}. The birth cluster of the Sun therefore has long since been dissipated, and its members, the solar siblings, are scattered throughout the Milky Way \citep[e.g.,][]{Bland-Hawthorn-10}. 

Finding solar siblings is important for several reasons. It would help us to better understand the origin of our Sun and to constrain its birthplace and the environmental conditions of the Sun's birth cluster. Moreover, finding and characterizing planetary systems (e.g., their frequency and architecture) around solar siblings could give relevant information about the outcome of planet formation in a common environment. Solar siblings can also be good candidates to search for planets with life, assuming that the life transportation between solar systems in the Sun's birth cluster was efficient \citep[e.g.,][]{Adams-05, Tepfer-06}. The highly speculative hypothesis that life (biotic materials) can travel in space and can settle in new habitats is called  panspermia. The transfer of life between exoplanet systems in particular is called interstellar lithopanspermia \citep{vonBloh-03}.

Several attempts have been made to find solar siblings \citep[e.g.,][]{Brown-10, Bobylev-11, Liu-15}, but only four plausible candidates have been identified to date: HIP21158, HIP87382, HIP47399, and HIP92831 \citep{Batista-12, Batista-14, Ramirez-14}. In most of the solar sibling search studies, the authors started their search from a kinematic selection of  candidates and then verified if th metallicities and chemical abundances of the candidates are compatible with those of the Sun. 

Following a different approach, \citet{Batista-14} conducted a search for solar siblings   in the HARPS high-resolution FGK dwarf sample \citep{Adibekyan-12} using a new approach based on the observed chemical abundance trends with condensation temperature. The initial HARPS sample used by these authors consisted of 1,111 stars, the abundances of only 12 elements were used to search for solar chemical twins, and finally, astrometric data were taken from Hipparcos \citep{vanLeeuwen-07}. \citet{Batista-14} found only one candidate solar sibling.  Today, we have the possibility to make a new search with improvements in all these aspects: a larger sample, chemical abundances of many more elements, and astrometric data with better precision. All these possibilities are provided by the AMBRE project \citep{deLaverny-13} and Gaia DR2 \citep{Gaia-18}. AMBRE is a Galactic archeology project set up by ESO and the Observatoire de la C\^{o}te d’Azur in order to determine the stellar atmospheric parameters for the archived spectra from the ESO spectrographs FEROS, HARPS, UVES, and GIRAFFE. A total of about 230,000 spectra have been homogeneously analyzed. 
  
This paper is organized as follows. In Sect.~\ref{init_samp_select} we describe the initial selection of the sample stars, and in Sect.~\ref{properties} we characterize these
stars in terms of stellar parameters (Sect.~\ref{param}), chemical abundances (Sect.~\ref{abund}), ages and activities (Sect.~\ref{ages}), and kinematics (Sect.~\ref{kinematics}).
In Sect.~\ref{best_candidates} we apply different criteria to select the best-fit solar sibling candidates, which we study in more detail in Sect.~\ref{best_4}. We summarize our work in Sect.~\ref{discussion}.

\section{Initial sample selection} \label{init_samp_select}

Our initial sample is based on AMBRE project data \citep{deLaverny-13}. Currently, the AMBRE project provides stellar parameters and the chemical index [$\alpha$/Fe] for 6,508 FEROS archived spectra \citep{Worley-12}, 10,212 UVES spectra \citep{Worley-16}, and 93,116 HARPS spectra \citep{DePascale-14}. These spectra correspond to about 17,000 individual stars. The spectra were processed using the MATISSE algorithm \citep{Recio-Blanco-06}, and we used a specific grid of synthetic spectra assembled by \citet{deLaverny-12}.
  
From this initial AMBRE sample we selected 28,631 spectra for which the MATISSE parameterization suggested a mean metallicity ([M/H]) and $\alpha$-element abundance [$\alpha$/H]) close to the solar value within  ($\pm$0.1 dex). These spectra correspond to 1,019 unique stars, for 987 of which Gaia DR2 \citep{Gaia-18} astrometry is available. 

The numerical simulations of \citet{Martinez-Barbosa-16} for solar sibling candidates with a parallax $\varpi$ $>$ 5 mas suggest a proper motion ($\mu$)  in the range 4 $< \mu < $  6 mas yr$^{-1}$ and a heliocentric radial velocity (RV) in the range $-2 <$ RV $< 0$ km/s. We note that these kinematic criteria are only suggestive and are used to increase the probability of finding a solar sibling. Moreover, they are dependent on which model of the Milky Way is assumed.  We decided to apply very broad initial kinematic criteria to avoid excluding any potential candidate at this stage. From the 987 stars we selected 119 relatively nearby stars ($\varpi$ $>$ 5 mas) with an RV between $-100$ and 100 km/s, and total $\mu < 50$ mas/yr. Following \citet{Lindegren-18}, we corrected the parallaxes for the global offset (0.029 mas) and for the $\sim$ 30\% underestimation of the parallax uncertainties of bright stars \citep{Luri-18, Arenou-18}. We did not consider distant stars with $\varpi$ $<$ 5 mas (about 10\% of the sample stars) because predictions of the kinematic properties of solar siblings are provided for nearby stars. Moreover, most of these distant stars are either very faint (the average V magnitude is $\sim$ 11.8 mag) or evolved massive giant stars (M $>$ 1.3 M$_\odot$). The spectra of faint stars are also usually of low quality, and it is difficult to derive precise chemical abundances. The massive evolved stars have lifetimes shorter than the present-day age of the Sun.

We finally made an intensive literature search for the 119 selected stars and excluded 26 stars belonging to open clusters that are much younger than the Sun (younger than 1 Gyr). Three stars with effective temperature (\teff) lower than 4500 K were also excluded since our spectroscopic analysis method does not guarantee a derivation of very precise stellar parameters and chemical abundances (at the level of $\lesssim$ 0.1 dex) for these very cool stars. After these selections, we had a sample of 90 stars. We study them in detail in the next sections.
  
\section{Properties of the sample stars} \label{properties}

\subsection{Stellar atmospheric parameters} \label{param}

In this section we describe how we derived precise stellar parameters for the selected 90 stars for which the initial parameterization by MATISSE suggests that they are solar sibling candidates. For most of the selected stars, several high-resolution (R$\sim$100.000) spectra were publicly available at the ESO archive. In order to work with spectroscopic data of the highest possible quality, we coadded all the available spectra of the stars that were observed with the same instrument and the same setup. For a few stars, spectra obtained with different instruments were available, and we selected spectra with the highest signal-to-noise ratio (S/N). In this work the signal-to-noise ratio refers to an S/N per angstrom at around 6000 \AA{}. For one of the stars (HD5418), an UVES spectrum was available that covers only blue wavelengths. This wavelength coverage was insufficient to derive the parameters with our technique, and this star was thus excluded from the sample.
  
The coadded spectra were used to derive precise stellar atmospheric parameters (\teff, [Fe/H], $\log g$, and \vtur) for the target stars, similar as in our previous works \citep[][]{Sousa-08,Sousa-15a}. For a detailed description we refer to \citet{Sousa-14}. The method is based on classical curve-of-growth analysis where the equivalent widths (EW) of the spectral lines are automatically measured with the ARES v2 code\footnote{The last version of the ARES code (ARES v2) can be downloaded from http://www.astro.up.pt/$\sim$sousasag/ares} \citep{Sousa-15}. Under the assumption of local thermodynamic equilibrium (LTE), the parameters were derived by imposing excitation equilibrium and ionization balance for \ion{Fe}{I} and \ion{Fe}{II} lines. We used the grid of ATLAS9 plane-parallel model atmospheres \citep{Kurucz-93} and the 2014 version of the MOOG\footnote{The source code of MOOG can be downloaded from http://www.as.utexas.edu/$\sim$chris/moog.html} radiative transfer code \citep{Sneden-73}. For stars with \teff $<$ 5200 K, we derived the parameters using the line-list compiled in \citet{Tsantaki-13}. This list represents a selection of lines presented in \citet{Sousa-08} that do not suffer from blends at low temperatures.   

\begin{figure}
\begin{center}
\includegraphics[width=1.0\linewidth]{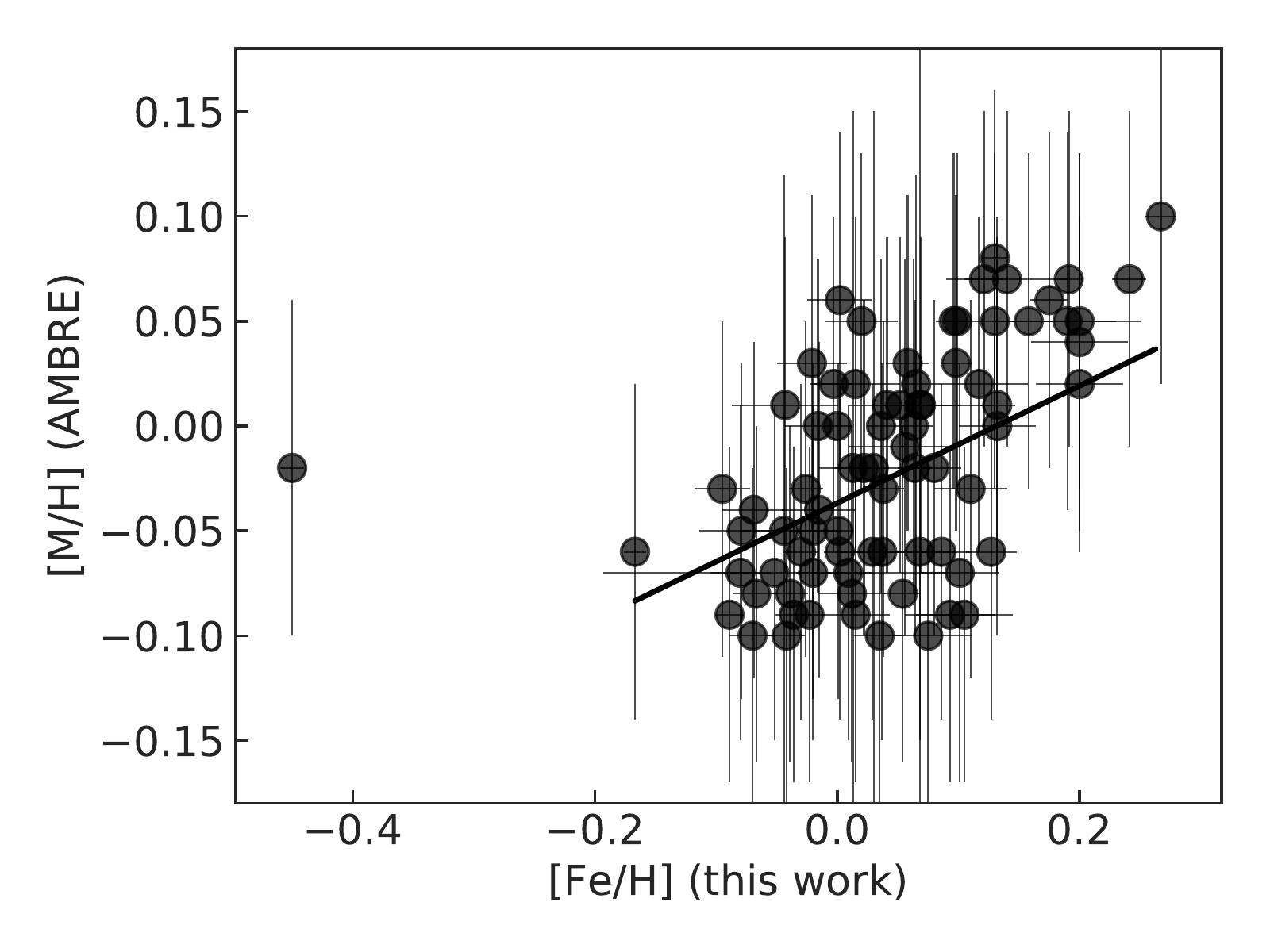}
\vspace{-1.cm}
\end{center}
\caption{Comparison between the AMBRE stellar mean metallicity and [Fe/H] derived in this work from spectra with higher S/N. The black solid line shows the result of the weighted least-squares regression. The star with the lowest [Fe/H] is not included in the fit.}
\label{fig-feh_main}
\end{figure}

\begin{table}[]
\caption{\label{tab:CPD-60315} Main properties of the binary system CPD-60315A and CPD-60315B.}
\centering
\small
\begin{tabular}{lll}
\hline\hline
        Star &      CPD-60315A &     CPD-60315B \\
\hline
Gaia DR2 & 4677970179987090304 & 4677970179988001280 \\
Gmag (mag) & 9.8541$\pm$0.0005 & 9.6830$\pm$0.0006 \\
$\varpi$ (mas) & 9.5809$\pm$0.0277 & 9.5838$\pm$0.0271 \\
RV (km/s) & 11.9291 &  11.5168 \\
\hline
\end{tabular}
\end{table}

For eight stars (HD217738, HD207889, HD217739, HD13021, TYC 964-160-1, BD-114934B, BD-114934B, and BD-202665A), the combined spectra had an S/N $<$ 50. The spectroscopic analysis of these low-S/N spectra did not allow us to derive very precise stellar parameters and chemical abundances. Our analysis of these stars suggested that they are all more metallic than the Sun by at least 0.1 dex, that is, [Fe/H] $>$ 0.1 dex. These stars were excluded from the sample. Moreover, the spectroscopic analysis showed that HD22556 is a spectral binary, and it was excluded from the sample. HD66488 and HD110108 were also excluded since we were unable to derive precise parameters (the method did not converge) for these hot stars (\teff $>$ 6500 K). Finally, we decided to exclude HD29167, which is a member of a binary system that consists of two practically identical stars: CPD-60315A and CPD-60315B (see Table~\ref{tab:CPD-60315}). The Gaia DR2 astrometry suggests that they have very similar parallaxes (the difference is 0.0029 mas, while the error of parallax for each component is at least ten times larger), and \textit{G} magnitudes (the difference is 0.17 mag). ESO archive contains two HARPS spectra for each of the components with an average S/N $\approx$ 30. The spectra are almost identical and suggest almost the same RV. It is unclear if the individual spectra obtained for each component may be contaminated by the companion. We decided to exclude this system from our sample. However, it is a very interesting system to be studied in more detail using spectra with higher quality and ensuring that the two components are observed separately.
  
In Fig.~\ref{fig-feh_main} we show the comparison between the mean stellar metallicity ([M/H]) derived by the AMBRE team and the metallicity derived in the current work based on iron lines alone ([Fe/H]). The plot shows that in addition to an offset of 0.06 dex and a scatter of 0.09 dex, there is a trend: the overestimation of [Fe/H] relative to [M/H] is higher for stars tht are more metallic. This probably reflects the influence of chemical elements other than iron in the definition of the [M/H] index. HD171028\footnote{This star also belongs to the HARPS metal-poor sample of \citet{Santos-11} and is known to host a giant planet \citep{Santos-07}.} is clearly an outlier in this plot. Our literature search confirms that it is a metal-poor star, while AMBRE parameterization resulted in a near solar metallicity. The star has 79 spectra in the ESO archive, 47 of which are parameterized by AMBRE. The  AMBRE mean metallicity for this star (excluding the results obtained for the 'outlier' spectrum) is $\sim-$0.64, which is even lower than what we report in this work. The same happened for some other stars that showed the most discrepant results between AMBRE and this work. It is important to recall that the AMBRE metallicities are derived for individual spectra (sometimes of low quality), while the [Fe/H] in this work is derived from the combined spectra. Without HD171028, the mean difference between [M/H] and [Fe/H] is 0.06$\pm$0.07 dex. The errors of the mean difference here represent the scatter, that is, standard deviation of the differences. To derive the standard error of the mean, these values need to be divided by the square root of the sample size. We note that the average error for [M/H] is 0.09 dex and for [Fe/H] is 0.03 dex. Thus the obtained difference is within the errors of our metallicity derivations.

A similar difference (0.072$\pm$0.10 dex) between AMBRE [M/H] and [Fe/H] (derived by spectral synthesis method) was observed in \citet{Mikolaitis-17} for a much wider range of stellar metallicities. These authors also stressed that these two parameters are not exactly the same and can be different for stars that have different element-over-iron abundances when compared to the solar values. In order to understand the possible reasons of the observed discrepancies, we show in Fig.~\ref{fig_feh_ambre_thiswork} the [Fe/M] ratio against the stellar atmospheric parameters, and parameters that might influence the results (e.g., S/N, or the CCF FWHM as a proxy of rotational velocity). This ratio, in addition to metallicity, correlates with the microturbulent velocity, effective temperature, and CCF FWHM (full-width at half-maximum of the cross-correlation function). In addition, we also derived the stellar parameters with our EW method using the individual spectra of the stars that were used in the AMBRE project. The comparison between the metallicities derived from individual and coadded spectra is shown in Fig.~\ref{fig_slope_atmos_param}. The plot shows a very good agreement, which suggests that the slightly discrepant results obtained in the AMBRE project and this work are probably related to the methods with which the stellar parameters are derived.
   
Similar to stellar metallicity, in Figs.~\ref{fig_logg_ambre_thiswork} and ~\ref{fig_teff_ambre_thiswork} we compare the surface gravity and \teff \ derived by the AMBRE team and in this work. For \logg, \ we observe a negligible offset of 0.03 dex and a scatter of 0.22 dex. This scatter can be explained with the average errors of 0.25 dex and 0.06 dex reported for AMBRE and in this work, respectively. For the \teff \ , we observe a good agreement between the two works. The 41$\pm$91 K difference is within the reported errors of 96 and 33 K for AMBRE and this work, respectively. 

For the further search of the best-fit solar sibling candidates, we restricted our final sample to the stars with $-0.1 \leqslant$ [Fe/H] $\leqslant$ 0.1 dex, which left us with a sample of 55 stars. The stellar parameters of these stars are presented in Table ~\ref{tab:params}. We note that only 23 of these stars were included in the search for solar sibling by \citet{Batista-14}.
  
\subsection{Trigonometric surface gravity} \label{trigonometric}
  
A known difficulty of spectroscopic methods is that they cannot constrain the stellar surface gravity with good accuracy  \citep[it is typically lower than $\sim$0.2~dex, e.g.,][]{Sozzetti-07, Kordopatis-11, Mortier-13, Tsantaki-14}. It has been shown that the spectroscopic surface gravity derived with the assumption of an LTE ionization balance does not always agree well with the \logg \ derived from the transit light curves of exoplanet host stars and asteroseismology \citep[e.g.,][]{Mortier-14}. A clear disagreement is also observed between spectroscopic and trigonometric surface gravities \citep[e.g.,][]{Tsantaki-13, Bensby-14, Delgado-Mena-17}. This observed difference between spectroscopic and trigonometric \logg \ depends on the \teff \ of the stars \citep[e.g.,][]{Delgado-Mena-17}. The trigonometric \logg \ values probably represent the real surface gravity of the stars better since they follow the isochrones in the Hertzsprung-Russell (HR) diagram far better than the spectroscopically derived ones \citep[see e.g.,][]{Delgado-Mena-17}. Thus we decided to derive trigonometric \logg \ for the sample stars using Gaia DR2 astrometric data. We note that if no trigonometric \logg \ is available, the corrections of spectroscopic surface gravities proposed by \citet{Mortier-14} can be used. Since all the targets are bright stars from the solar vicinity, Gaia provides very accurate parallaxes. Only four stars out of the 55 have relative error in $\varpi$ worse than 2\%, with the largest error being 6.4\%.
 
\begin{figure}
\begin{center}
\includegraphics[width=1\linewidth]{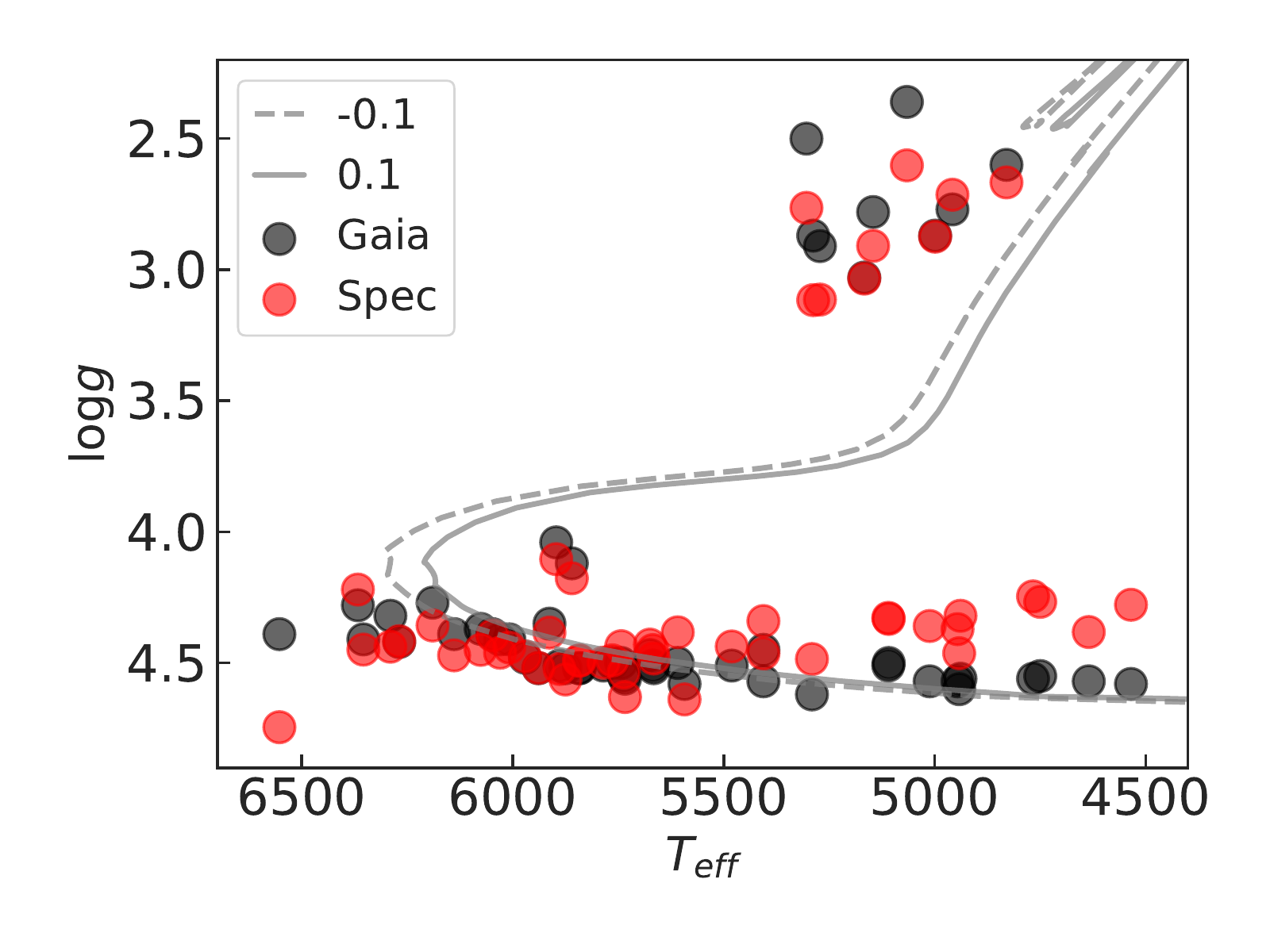}
\vspace{-1.cm}
\end{center}
\caption{HR diagram for the sample stars with the spectroscopic (Spec) and trigonometric (Gaia) surface gravities. The solar age isochrones for stars with metallicity $-$0.1  and 0.1 dex are shown in gray.}
\label{fig-hr}
\end{figure}

\begin{figure*}
\begin{center}
\includegraphics[width=0.7\linewidth]{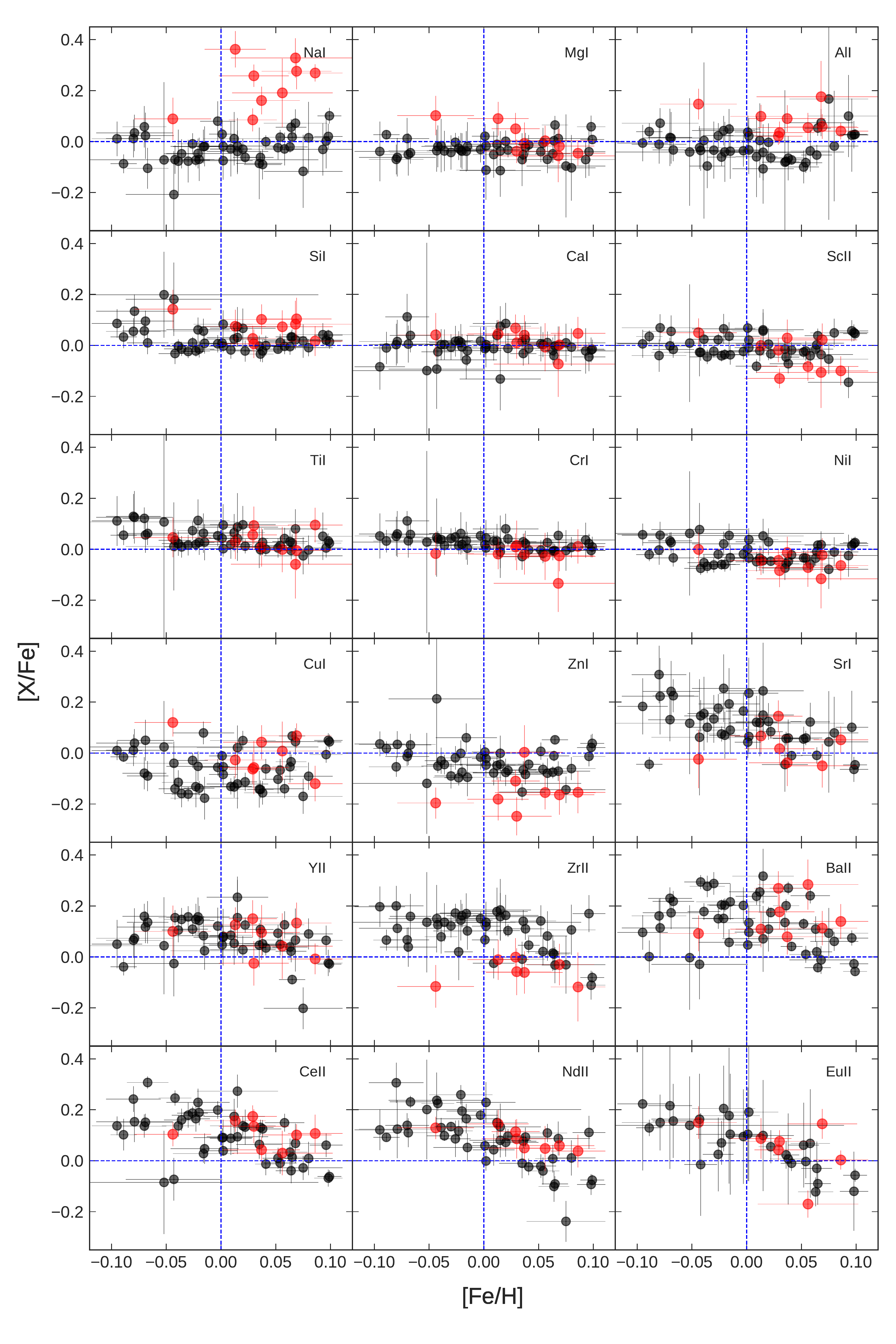}
\vspace{-0.8cm}
\end{center}
\caption{Abundance ratio for a range of species X, [X/Fe] against stellar metallicity for the current sample. The evolved (and also massive) stars with surface gravity $<$ 3.5 dex are shown in red, and the stars at earlier stages of their evolution (\logg \ $>$ 3.5 dex) are shown in black. The blue dashed horizontal and vertical lines show the solar abundances.}
\label{fig-elfe_feh}
\end{figure*}

We calculated the trigonometric \logg \ using the well-known Newton law of universal gravitation and the Stefan–Boltzmann law \citep[see, e.g., Eq. 1 from][]{Santos-04}. We used the Gaia DR2 parallaxes \citep{Gaia-18}, V magnitudes extracted from Simbad\footnote{http://simbad.u-strasbg.fr/simbad/}, bolometric corrections based on \citet{Flower-96}, and revisited by \citet{Torres-10}, the stellar masses and spectroscopic \teff. Stellar masses, together with the radii and ages of the stars, were derived from the PARAM v1.3 web interface\footnote{http://stev.oapd.inaf.it/cgi-bin/param} based on the Padova theoretical isochrones from \citet{Bressan-12} and with the use of a Bayesian estimation method \citep[][]{daSilva-06}. As input parameters for PARAM, we used the V mag, $\varpi$, and spectroscopic \teff \ and [Fe/H]. As priors for the mass, the initial mass function of \citet{Chabrier-01} was used. No correction for interstellar reddening was needed since all the stars are nearby objects within 200 pc. The results are presented in Table~\ref{tab:params}. 

In Fig.\ref{fig-hr} we show the spectroscopic HR diagram for the sample stars together with the Padova isochrones for solar age and metallicity. The plot clearly demonstrates that the spectroscopic \logg \ values do not follow the expected stellar evolution curves for cool dwarfs. It is also clear that the evolved stars are too far away from the solar age isochrones (they are mostly very young: see Sect. ~\ref{ages}). 

\subsection{Chemical abundances} \label{abund}
 
When we derived chemical abundances of the elements, we generated the model atmospheres for the spectroscopically derived \teff \ and [Fe/H], and the trigonometric \logg. We did not rederive the \teff \ and [Fe/H] by fixing the \logg \ since the effect of an unconstrained \logg \ on the derivation of other stellar parameters is small for the EW-based curve-of-growth approach \citep{Torres-12, Mortier-13}. Moreover, although the effect is not very strong, fixing \logg \ can bias the results and derivation of other atmospheric parameters \citep{Mortier-14, Smalley-14}. 

The chemical abundances were derived in the same way as in our previous works \citep[e.g.,][]{Adibekyan-12, Delgado-Mena-17}. We used the same tools and models of atmospheres as for the derivation of stellar parameters. For elements with only a few spectral lines, the EW measurements of ARES were made with careful visual inspection of the spectra. In some very few cases, when the ARES measurements were obviously incorrect (e.g., as a result of cosmic rays or bad pixels close to the spectral line), we manually measured the EWs using the task \texttt{splot} in IRAF\footnote{IRAF is distributed by National Optical Astronomy Observatories, operated by the Association of Universities for Research in Astronomy, Inc., under contract with the National Science Foundation, USA.}. The final abundances of the elements with more than three measured spectral lines were calculated as a weighted mean of the estimates from each line, where the inverse of the distance from the median abundance was considered as a weight. This method can be effectively used without removing suspected outlier lines \citep{Adibekyan-15}. For the elements with only two or three lines, we calculated the errors on EWs following \citet{Cayrel-88} to provide more realistic errors for the abundances of elements. This uncertainty takes into account the statistical photometric error that is due to the noise in each pixel and the error related to the continuum placement. The latter contributes most to the total error \citep[][]{Cayrel-88, Bertran-15}. These errors were then propagated to derive the abundance uncertainties for each spectral line. The final uncertainties for the average abundance were propagated from the individual errors. The final errors of the [X/H] abundances were calculated as a quadratic sum of the errors due to EW measurements and errors due to uncertainties in the atmospheric parameters. The solar reference abundances were taken from \citet{Adibekyan-16} and \citet{Delgado-Mena-17}, which were derived using the combined HARPS reflected spectrum (S/N $\sim$ 1300) from Vesta, which we extracted from the ESO public archive. We note that the choice of asteroids from which to obtain the solar reflected spectra and observations at different epochs have effects smaller than $<$ 0.01 dex \citep[e.g.,][]{Bedell-14}.

In Figs.~\ref{fig-elfe_feh} and ~\ref{fig_elfe_logg_teff} we show the dependence of [X/Fe] abundance ratios on the metallicity, trigonometric surface gravity, and effective temperature. The massive and evolved stars (\logg \ $<$ 3.5 dex and M $>$ 1.5 M$_\odot$) of the sample are shown in red and the dwarf solar-type stars are in black. The abundance of all the stars is presented in an electronically readable table at the CDS. We note that for two relatively fast rotators (HD75006 and HD62816) we were unable to derive abundances of all the elements.

\subsection{Stellar ages and activity} \label{ages}

\begin{figure}
\begin{center}
\begin{tabular}{c}
\includegraphics[angle=0,width=0.9\linewidth]{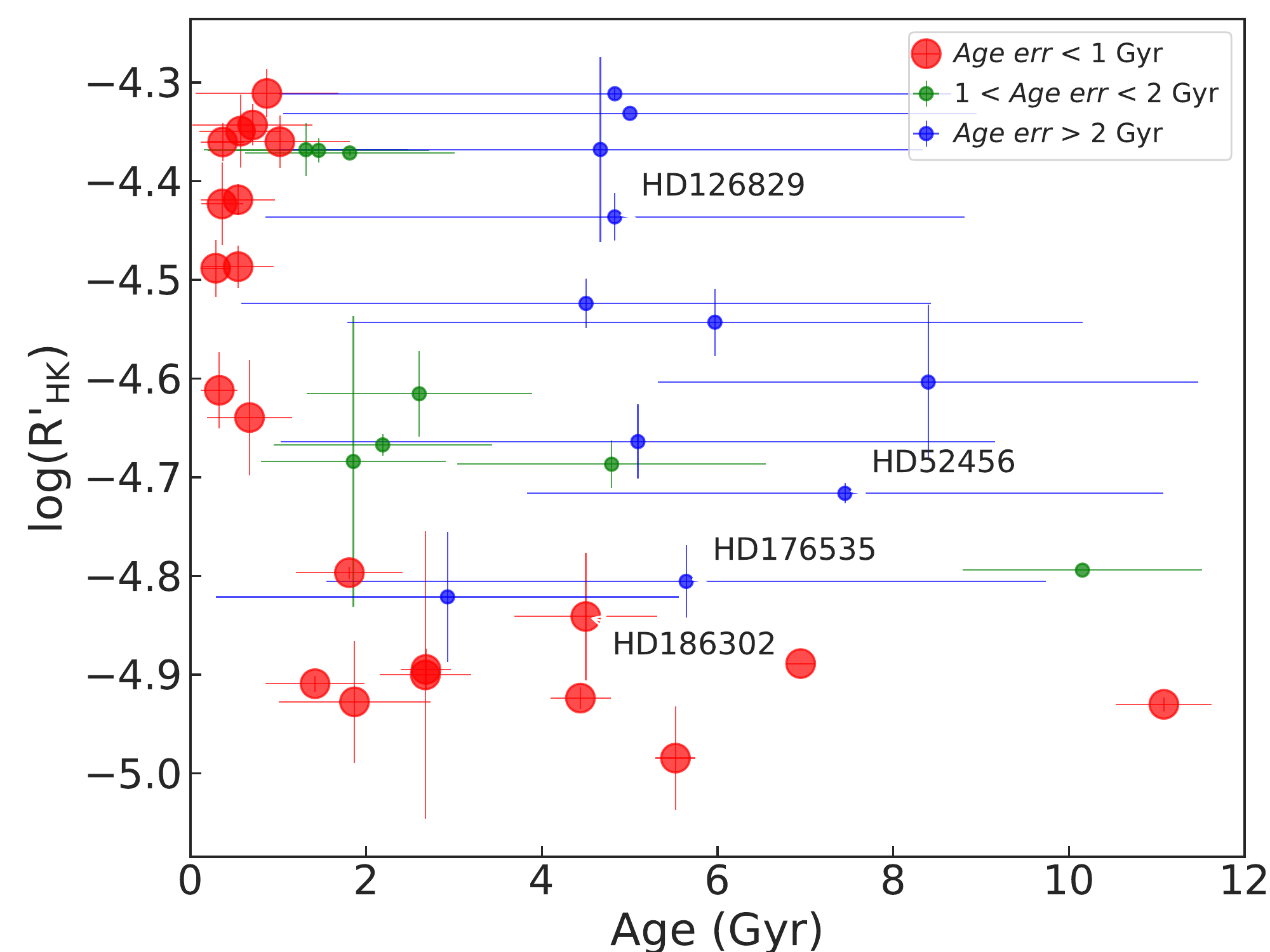}
\vspace{-0.4cm}
\end{tabular}
\end{center}
\caption{Stellar activity index $\log$(R$^{'}_{\mathrm{HK}}$) against ages for the stars with masses lower than 1.2 M$_{\odot}$. The names of the four best-fit candidates are indicated in the plots.}
\label{fig-activity_age}
\end{figure}

One of the main parameters for identifying solar siblings is the stellar age, as the solar siblings are expected to share the same age. As discussed in Sec.~\ref{trigonometric}, the stellar ages for the sample stars were derived using the Padova isochrones (Table~\ref{tab:age}). Although the precise astrometric data of Gaia help increase the precision of stellar ages, for a significant fraction of the stars, the uncertainties of the age estimations are unfortunately still $>$ 2-3 Gyr. 

Several stellar parameters correlate with stellar age. These chronometers can be used, at least, to cross-check the isochronal ages of the stars. One of the most frequently discussed chronometers is the $\log$(R$^{'}_{\mathrm{HK}}$) chromospheric activity indicator \citep[e.g.,][]{Mamajek-08, Pace-13, Lorenzo-Oliveira-16, Lorenzo-Oliveira-18}. We followed the works of \citet{Suarez-15} and Hojjatpanah et al. (2018, in prep.) to measure the emission flux in the cores of the \ion{Ca}{II} H\&K lines and determine the activity index. For most of the stars we had HARPS spectra available, which cover the wavelength region of \ion{Ca}{II} lines. However, for two stars (HD209458 and HD92987), high-resolution UVES spectra that were used to spectroscopically characterize the stars did not cover the \ion{Ca}{II} lines. To determine the activity indexes of these stars, we used the HARPS-N\footnote{HARPS-N is the northern copy of HARPS and has the same resolution and same spectral coverage \citep{Cosentino-12}.} spectrum for HD209458 and the R = 58,000 resolution UVES spectrum that covered 3732-4999\AA{} region for HD92987. The activity indices of the stars can be found in Table ~\ref{tab:age}.

\begin{figure*}
\begin{center}
\begin{tabular}{ccc}
\includegraphics[angle=0,width=0.45\linewidth]{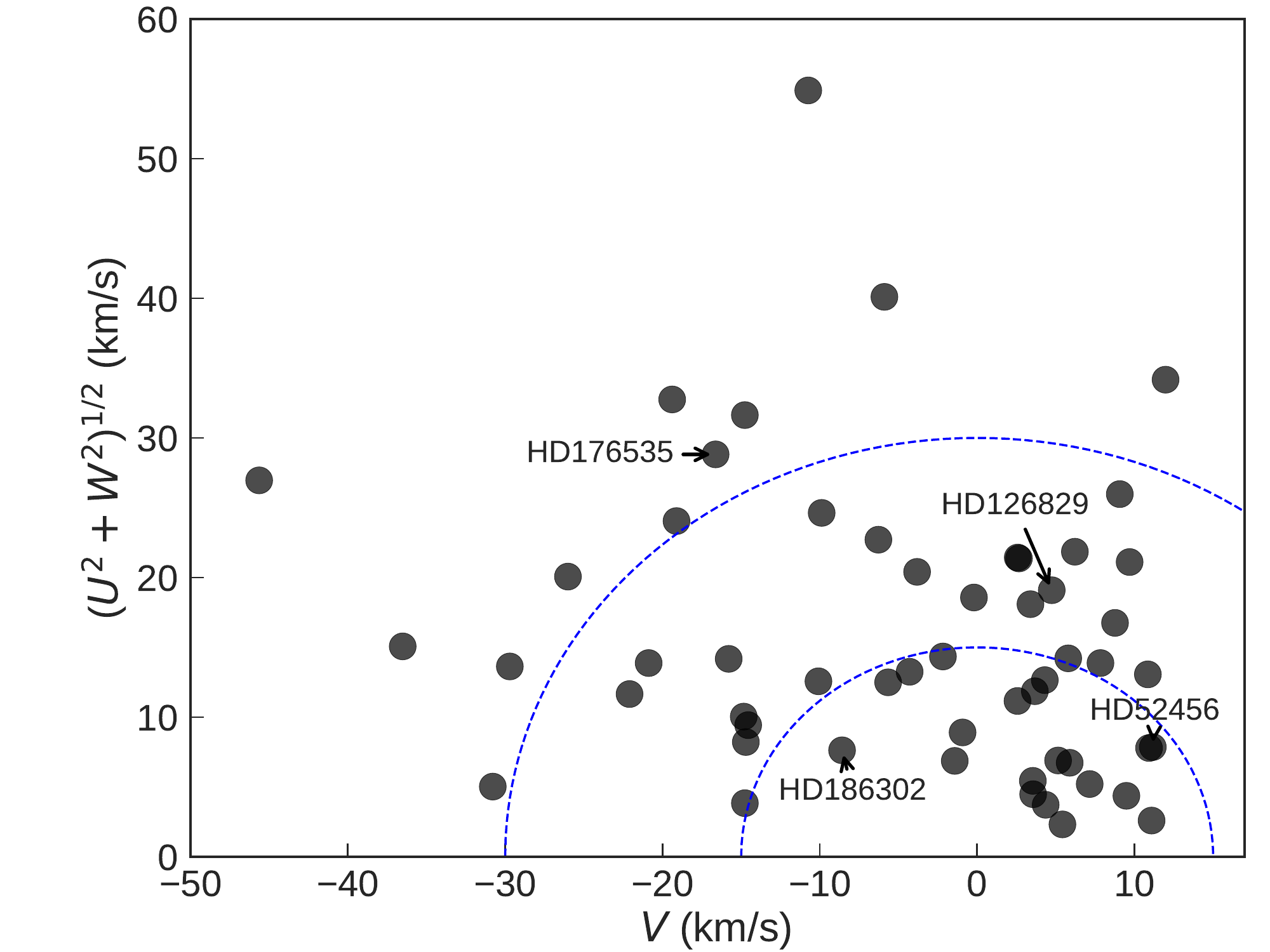}
\includegraphics[angle=0,width=0.45\linewidth]{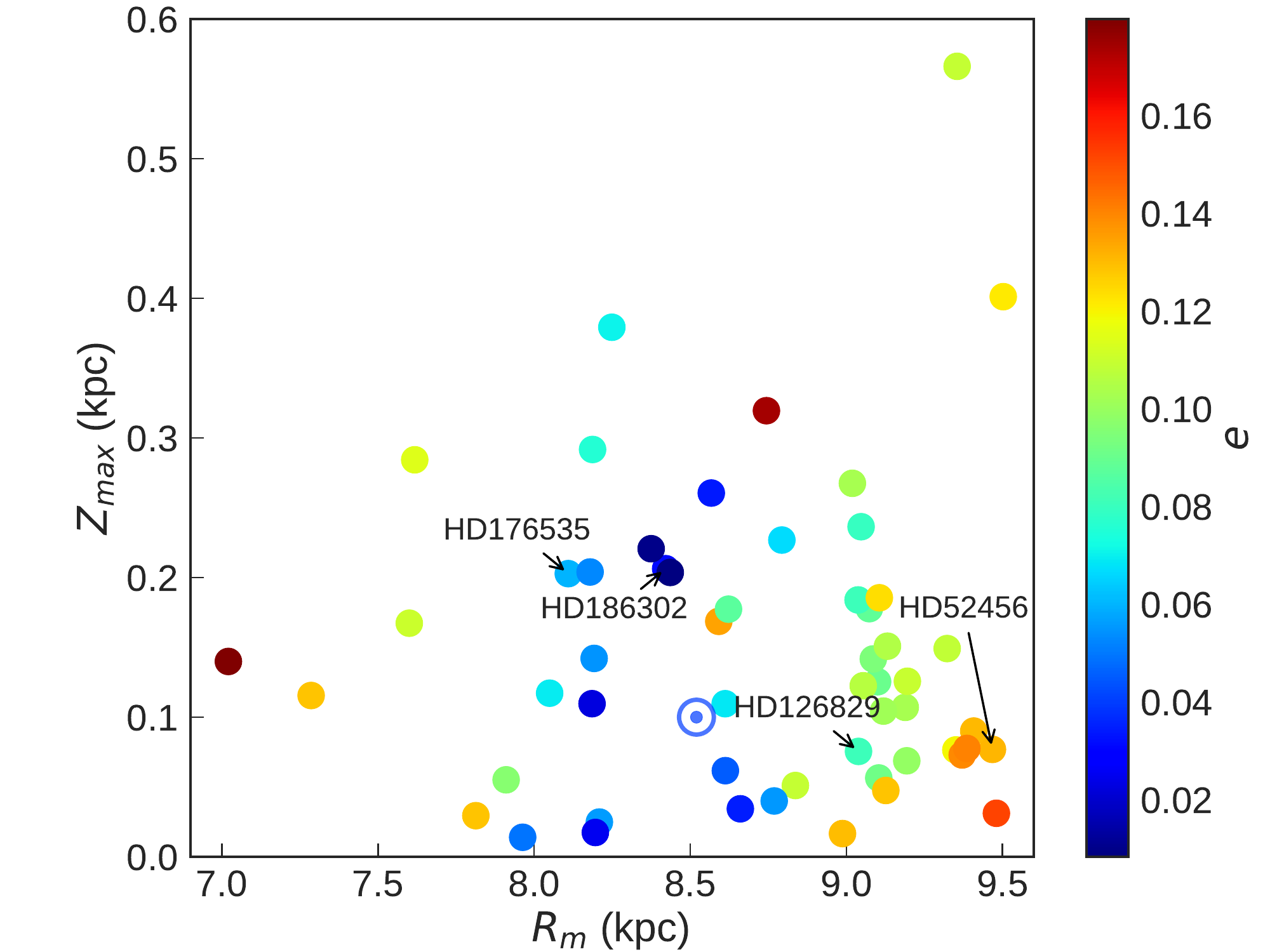}
\vspace{-0.4cm}
\end{tabular}\end{center}
\caption{\textit{Left panel:} Distribution of the velocities of our sample of solar sibling candidates in the Toomre diagram. \textit{Right panel:} Z$_{max}$ (maximum height from the Galactic plane) vs. R$_{m}$ (mean orbital radius) for the sample stars. The color bar corresponds to the orbital eccentricities of the stars. The position of the Sun is marked by the solar symbol. The four best-fit solar sibling candidates are indicated by their names in the plots.}
\label{fig-kinematics}
\end{figure*} 
  
In Fig.~\ref{fig-activity_age} we show the dependence of stellar activity on age for the stars with masses $<$ 1.2 M$_{\odot}$. This mass limit corresponds to $\sim$F3 spectral type. Stars of spectral types earlier than $\sim$F4 have shallower convection zones and show no correlation between stellar activity, rotation, and age \citep[e.g.,][]{Wolff-85, Garcia-Lopez-93}. Moreover, most of these massive stars are already evolved from the main sequence (MS), for which the age-activity relation is probably not valid \citep{Wright-04}. The figure clearly shows that all the active stars ($\log$(R$^{'}_{\mathrm{HK}}$) $\lesssim$ $-$4.7 dex) with age estimates with errors $<$ 1 Gyr are younger than about 1.5 Gyr. The group of active stars that have isochronal ages older than 4 Gyr is also easily identified. These are the stars with the largest errors in age.

In addition to the chromospheric activity, some abundance ratios can also be used, such as [Y/Mg], \citep[e.g.,][Titarenko et al., 2018 submitted]{Nissen-15, TucciMaia-16, Slumstrup-17, Delgado-Mena-18}, and the Li abundance \citep[e.g.,][]{Sestito-05, Charbonnel-05, Andrassy-15} as an indicator of stellar age. We use these chronometers to verify the robustness of the derived ages for the best-fit solar sibling candidates in Sect.~\ref{best_candidates}. 
  
\subsection{Kinematics} \label{kinematics}
  
During its evolution, the Sun's birth cluster can have undergone many disruptive processes (e.g., dissolution of the cluster, and non-axisymetric perturbations due to the presence of the bar and spiral arms), which will have spread the solar siblings throughout the Milky Way. Several studies have provided predictions on the current distribution of the solar siblings \citep[e.g.,][]{Bland-Hawthorn-10, Brown-10, Bobylev-14, Martinez-Barbosa-16}. Depending on the assumptions about the Galactic model and the number of stars in the initial cluster, different numbers of solar siblings are suggested to be currently present in the solar neighborhood \citep[][]{Bland-Hawthorn-10, Martinez-Barbosa-16}. These solar siblings are also predicted to have a given kinematics \citep{Bobylev-11}, Galactic coordinates \citep[e.g.,][]{Bland-Hawthorn-10}, and  phase-space distribution \citep[][]{Martinez-Barbosa-16}. 

We calculated the Galactic space velocity components (UVW) by combining the Gaia astrometric data and the RV derived in the current work. When calculating the velocity components with respect to the local standard of rest (LSR), for the solar motion relative to the LSR we adopted the values of (U, V, W)$_{\odot}$ = (11.1, 12.24, 7.25) km s$^{-1}$ as measured by \citet{Schonrich-10}. The orbital parameters of the sample stars were also derived using the space velocities and the publicly available package {\texttt{galpy}}\footnote{ http://github.com/jobovy/galpy} \citep{Bovy-15}. We used the axisymmetric potential \texttt{MWPotential2014} \citep{Bovy-15}, and assumed R$_{\odot}$ = 8.34 kpc and Z$_{\odot}$ = 0.025 kpc \citep{Reid-14}. 

In the left panel of Fig.~\ref{fig-kinematics} we show the distribution of the stars in the Toomre diagram. The velocities of the stars are relative to the Sun. We also show the isovelocity curves of Vpec = (U$^{2}$ + V$^{2}$ + W$^{2}$)$^{1/2}$ = 15 km s$^{-1}$ and 30 km s$^{-1}$. The right panel of the same figure shows the dependence of the maximum height from the Galactic plane (Z$_{max}$) versus the mean orbital radius (R$_{m}$), color-coded by their eccentricities ($\it{e}$). The position of the Sun is also shown in the figure. We note that the kinematic parameters of the Sun were calculated using \texttt{galpy} following the instructions in the package manual. The kinematic properties of the stars are presented in Table ~\ref{tab:kinematics}.

\section{Selecting the best-fit solar sibling candidates} \label{best_candidates}
  
After deriving the main properties for the sample stars, we can select the best-fit solar sibling candidates. The two main and very strict conditions for solar siblings are i) that they need to be chemically similar to the Sun and ii) that they need to have the same age as the Sun.
  
\subsection{Siblings by chemistry}
  
The level of chemical similarity between the solar siblings depends on the chemical homogeneity of the Sun's birth cluster. Some theoretical works have suggested that open clusters are chemically homogeneous up to $\sim$10$^{5}$ $M_{\odot}$ in mass \citep[e.g.,][]{Bland-Hawthorn-10}. The very recent work by \citet{Armillotta-18} suggested that the spatial scale of the chemically homogeneous clusters consisting of about 2000 stars is $\sim$1 pc. The typical diameter of open clusters is a few parsec \citep{Schilbach-06}. The observed variation of chemical abundances in open clusters (star-to-star abundance scatter) that is due to turbulent mixing in star-forming molecular clouds is much smaller than the initial gas abundance scatter where the stars were formed \citep[e.g.,][]{Feng-14}. Most recent observational studies suggest that open clusters are homogeneous at about 0.03 dex level \citep[e.g.,][]{Hogg-16, Liu-16, Bovy-16}. 

In Fig.~\ref{fig_xh_tc_1} we show the chemical abundances of the stars as a function of atomic number. Some stars clearly show a systematic over- or underabundance in heavy elements (Z $>$ 30) when compared to the light elements. This high or low heavy-to-light element ratio relative to the solar value is probably a consequence of galactic chemical evolution. The interstellar medium becomes polluted in neutron-capture elements by stars at the asymptotic giant branch (AGB) phase. Thus stars with different ages might show different abundance ratios of light-to-light elements \citep[e.g.,][]{Melendez-14, Spina-16}. In addition, the enhancement of neutron-capture elements for an individual star can be due to pollution from a companion AGB star \citep{Liu-16a}.

To identify the solar sibling candidates, we selected the stars with metallicities and average chemical abundances of light and heavy elements close to the solar cluster value within the uncertainties. Because of the (possible) chemical inhomogeneity of the Sun's birth cluster described above, we assumed the solar cluster value for chemistry to be within 0.00$\pm$0.03 dex. The average chemical abundances of the light and heavy elements were calculated as the weighted mean using the inverse of the uncertainties of individual abundances as weights. For evolved massive stars we did not consider the abundance of Na when we calculated the mean abundances because it is overabundant when compared to the dwarfs \citep[see Fig.~\ref{fig_elfe_logg_teff} and e.g.,][]{Adibekyan-15a}. The overabundance of Na in giant stars is usually explained as a stellar evolutionary effect \citep[e.g.,][]{Jacobson-07}, although an inappropriate spectroscopic analysis method (e.g., not taking into account non-LTE effects, which are stronger for giants than for dwarfs \citep[][]{Alexeeva-14} can also affect the results. Note that the very recent work by \citet{Smiljanic-16} suggests that the Na abundance of stars with masses below $\sim$ 2 M$_{\odot}$ is not affected by internal mixing processes.
  
Our chemical and metallicity selection criteria left us with 12 stars:  HD2247, HD45415, HD52456, HD62412, HD74006, HD77191, HD99648,  HD115341, HD126829, HD176535, HD186302, and HD199951.

\subsection{Siblings by age}
  
\begin{table}[t!]
\caption{\label{tab:abund_age} Stellar ages based on different chemical abundance ratios for our best-fit sibling candidates.}
\centering
\small
\begin{tabular}{lccc}
\hline\hline
Star &       Age([Y/Mg])$^{1}$  &    Age([Y/Si])$^{1}$ &    Age([Y/Mg])$^{2}$\\
 & Gyr & Gyr & Gyr \\
\hline
HD186302 & 3.1$\pm$1.7 & 2.4$\pm$1.6 & 3.8$\pm$1.7\\
HD126829 & 3.5$\pm$3.3  & 9.4$\pm$5.1 & 4.2$\pm$3.4\\
HD176535* & -- &  7.8$\pm$4.5 & -- \\
HD52456 & 1.0$\pm$2.5 & 2.6$\pm$2.4 & 1.5$\pm$2.6\\
\hline
\end{tabular}
\noindent

$^{*}$No [Mg/H] abundances were derived for this star.\\
$^{1}$Based on Delgado Mena et al. (2018, in prep.). \\
$^{2}$Based on \citet{TucciMaia-16}.

\end{table}   

Very recently, \citet{Ness-18} demonstrated that at solar metallicity ([Fe/H] = 0.00$\pm$0.02 dex), 1\% of the field star pairs can have indistinguishable chemical abundances at a 0.03 dex precision level. Chemical identity to the Sun therefore is an important and mandatory criterion, but clearly not enough to identify stellar siblings. In this section we aim at selecting the candidates whose ages are compatible to the age of the Sun. 

Five of the 12 selected candidates (HD45415, HD62412, HD74006, HD99648, and HD199951) have  M $>$ 1.65 M$_{\odot}$. The lifetimes of these stars are therefore shorter than the current age of the Sun, and thus they cannot be solar siblings. Three other stars have isochronal ages shorter than 3 Gyr: 0.65$\pm$0.50, 1.46$\pm$1.26, and 2.67$\pm$0.52 Gyr for HD2247, HD77191, and HD115341, respectively. These three stars were thus excluded from the sample of best-fit candidates.

\begin{figure}
\begin{center}
\includegraphics[width=1\linewidth]{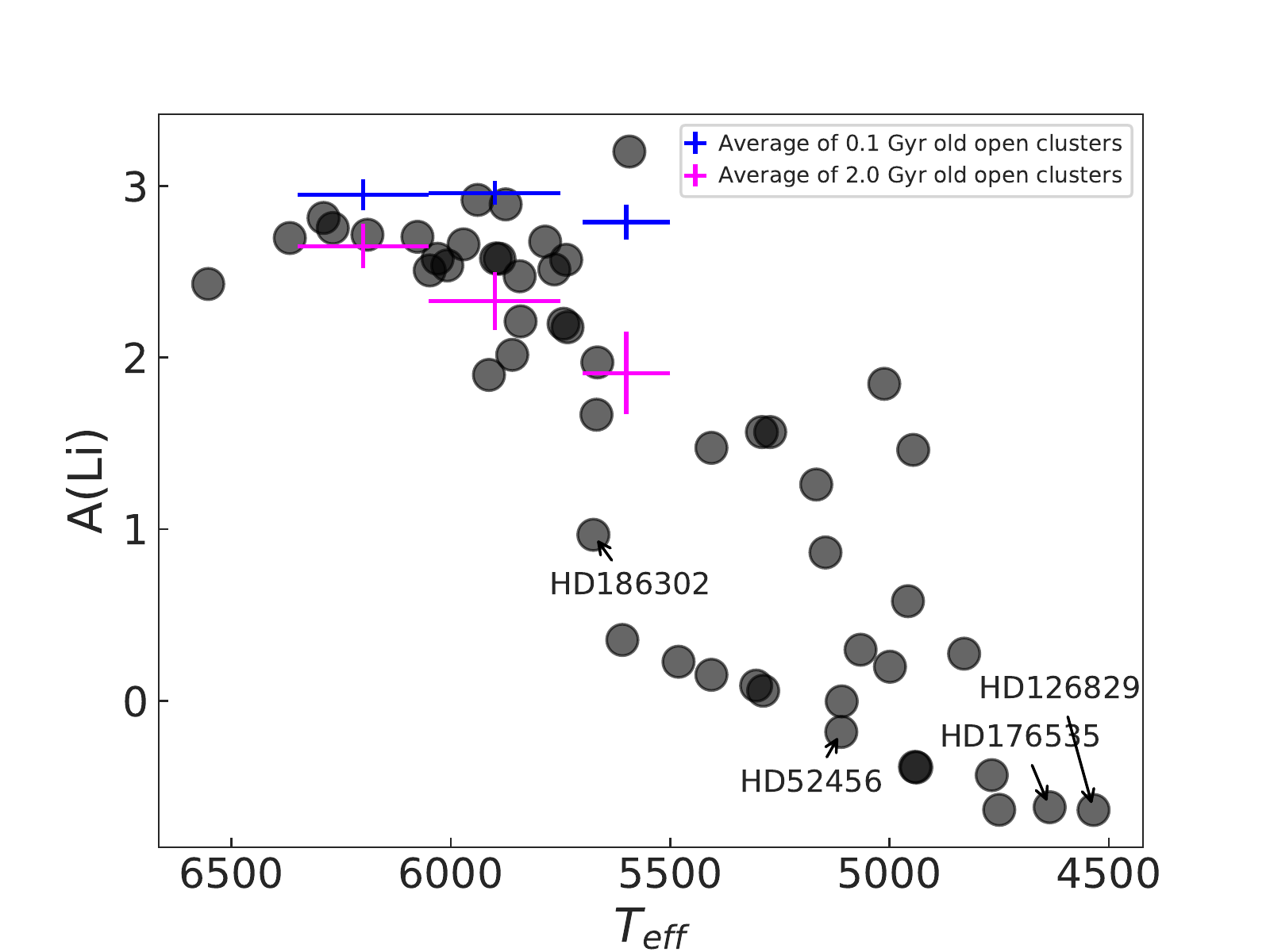}
\vspace{-0.8cm}
\end{center}
\caption{Li abundance vs. stellar effective temperature for the sample stars. The four best-fit candidates are indicated by their HD numbers. The average Li abundances for 0.1 Gyr and 2.0 Gyr open clusters (Pleiades and Blanco1, and Hyades, Praesepe, Coma Ber, and NGC 6633) in three ranges of \teff\ are also shown. The data are taken from\citet{Sestito-05}.}
\label{fig-li_teff}
\end{figure}

The remaining four stars (HD52456, HD126829, HD176535, and HD186302: see Table~\ref{tab:abund_age}) have isochronal ages similar to the age of the Sun within the uncertainties. It is important to note that except for HD186302, the uncertainties of isochronal ages are larger than 3.5 Gyr (see Table~\ref{tab:age}), which makes it very difficult to conclude whether they are solar siblings. One of these stars (HD126829) seems to be very active for a solar-age dwarf: $\log$(R$^{'}_{\mathrm{HK}}$) = $-$4.4 dex. The B--V color of this cool K-type star is beyond the range for which \citet{Mamajek-08} have calibrated their age-activity relation, and thus we cannot use this chronometric indicator. The relation between chromospheric activity and rotation (and age) is probably more complex at temperatures that are much cooler than the temperature of the Sun \citep[e.g.,][]{Noyes-84, Suarez-16}. However, we were able to use the age-activity relation of \citet{Mamajek-08} to derive the ages of HD186302 and HD52456, which are 3.8 and 2.2 Gyr, respectively. The errors on this age can be 30-60\%, depending on the activity level of the star \citep{Mamajek-08}.

As an independent chronometer, we also examined the Li abundance of these four stars. The Li abundances of the sample stars were derived from the EW measurements of the \ion{Li}{I} 6708\AA{} line. We note that to derive an accurate Li abundance, the spectral synthesis method is commonly used \citep[as we did previously, e.g.,][]{Delgado-Mena-14, Delgado-Mena-15}. However, for the purpose of verifying the stellar ages, the fast EW method is sufficient. In Fig.~\ref{fig-li_teff} we show the relation between Li abundance and \teff. The plot suggest that all four stars, including the active star HD126829, are not particularly young ($<$ 2 Gyr). 

Moreover, as we mentioned in Sect.~\ref{ages}, [Y/Mg] and [Y/Si] ratios can also be used to estimate the stellar ages. We used the relation between [Y/Mg] \ and age, and [Y/Mg] \ and age provided in Delgado Mena et al. (2018, in prep.),  and the [Y/Mg] \ versus age relation from \citet{TucciMaia-16}. The results are presented in Table~\ref{tab:abund_age}. Unfortunately, the errors of the age estimates are large and cannot be used to further constrain the ages. Nevertheless, the abundances of HD52456 suggest a young age ($<$ 3 Gyr) for this star. We note that the relations in Delgado Mena et al. (2018, in prep.) and \citet{TucciMaia-16} are obtained for solar-analog and solar-twin stars and do not cover the temperatures of the three coolest stars in our sample. Thus, the age estimates for these three stars derived by this method should be taken with more caution.
Unfortunately, the uncertainty of the age estimations based on the activity and chemical abundances are very large and do not help significantly in making a a better selection on these last candidates.

\section{Four best-fit sibling candidates} \label{best_4}
  
\begin{figure}
\begin{center}
\includegraphics[width=1\linewidth]{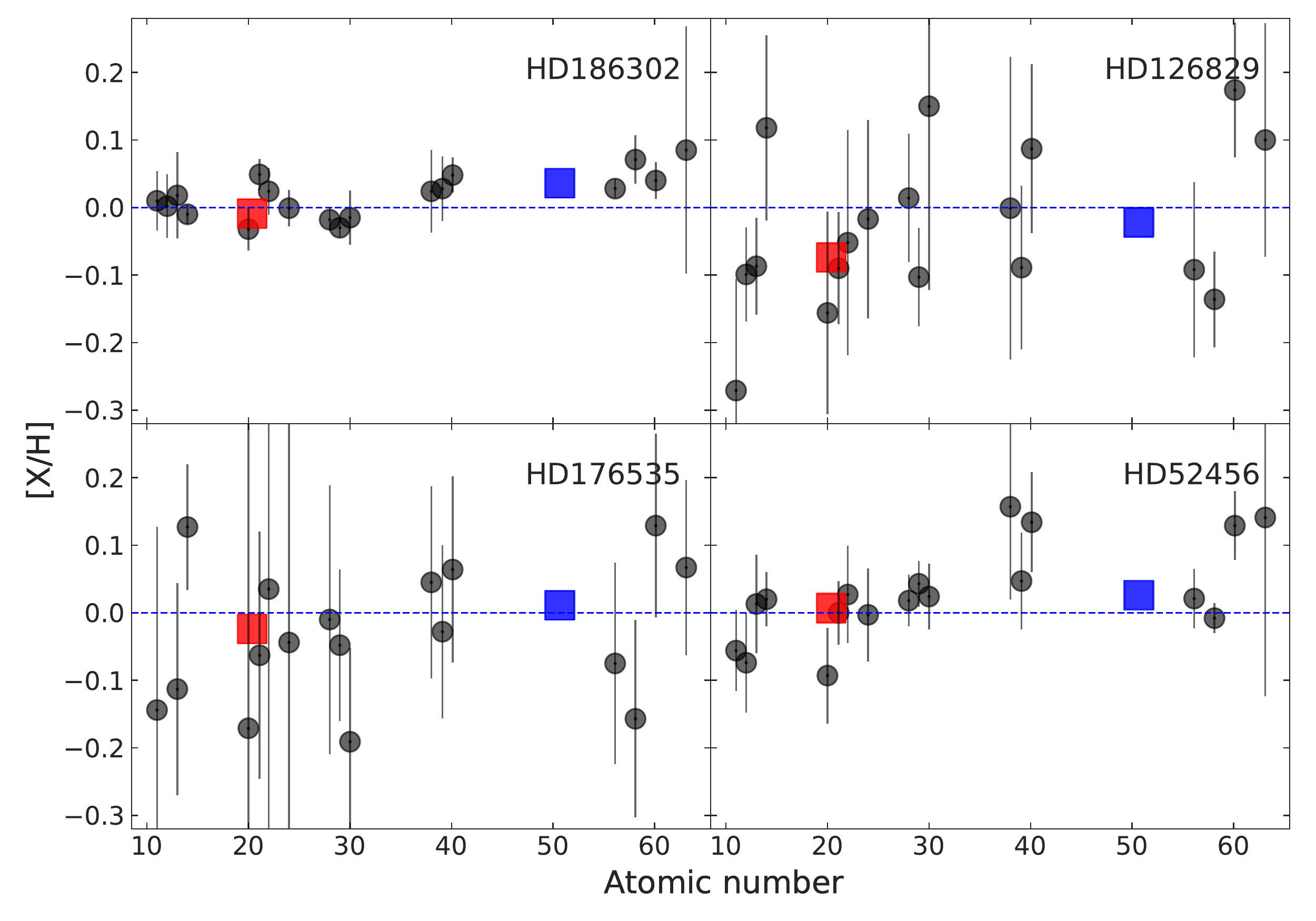}
\end{center}
\vspace{-0.4cm}
\caption{Abundances vs. atomic number for the four best-fit solar sibling candidates. The average abundances of light (Z $\leqslant$ 30) and heavy (Z $>$ 30) elements are shown by red and blue squares, respectively.}
\label{fig-xh_z_best}
\end{figure}  
 
In the previous sections we identified four solar sibling candidates whose chemical abundances (see Fig.~\ref{fig-xh_z_best}) and ages are similar to those of our Sun. In this section we explore the kinematic properties and the carbon isotopic ratio of these targets.
  
\subsection{Kinematics and orbits}
 
As described in Sect.~\ref{kinematics}, several attempts have been made to predict the present-day kinematics and orbits of the solar siblings. These predictions are highly dependent on the Galaxy model and the initial properties (e.g., mass and position in the Galaxy) of the parent cluster \citep[see, e.g.,][for details and discussion]{Bland-Hawthorn-10, Martinez-Barbosa-16}.

In terms of the peculiar velocity and orbital parameters, HD186302 seems to be the closest to the Sun (see Fig.~\ref{fig-kinematics}). HD176535 is the farthest star in terms of peculiar velocity and HD52456 is most different in terms of galactic orbital parameters. It is clear that these kinematic analyses are\ only very qualitative. For a quantitative kinematic study, the Galactic orbits of these candidates need to be construced and the orbits have to be integrated backward in time to determine whether their orbits encountered the orbit of the Sun about 4.6 Gyr ago. An obvious limitation of this technique is that serious assumptions are required for the Milky Way model.

\subsection{Carbon abundances and isotopic ratios}
  
We also tried to estimate LTE carbon abundances and carbon isotopic ratios ($^{12}$C/$^{13}$C) by spectral synthesis for our best-fit sibling candidates. The carbon isotopic ratio is indeed another good indicator of the initial composition of the parent cluster, since this ratio is expected to have remained the same in the atmosphere of dwarf stars (a strong decrease of $^{12}$C/$^{13}$C is only expected after the first dredge-up in low-mass stars). We also recall that isotopic ratios are insensitive to atmospheric parameter uncertainties because the lines originating from identical molecular excitation levels are compared. Moreover, the $^{12}$C/$^{13}$C ratio changes with galactocentric distance \citep[e.g.,][]{Wilson-92} and time \citep[e.g.,][]{Kobayashi-11} and, thus, this ratio could help to characterize the parent cluster. In summary, the solar sibling nature of our best-fit candidates would be supported if carbon abundances and $^{12}$C/$^{13}$C ratios that are close to solar were found.

For such abundance derivations, we analyzed the coadded HARPS spectra of our best-fit candidates around the G band where several C$^{12}$H and C$^{13}$H features are expected to be present. Very high resolution and high S/N spectra are required for such an analysis since the $^{13}$CH lines are  expected to be very weak in metal-rich cool dwarfs because most ($\sim$99\%) of the carbon is $^{12}$C. Moreover, a very good linelist is reqquired for the spectral synthesis since there are almost no isolated $^{13}$CH lines in such spectra, and several blends need to be investigated. Finally, the spectra of metal-rich cool dwarfs are also difficult to normalize in this spectral domain. Owing to such difficulties, no previous determinations of carbon isotopic ratios in metal-rich cool-dwarfs from CH lines are found in the literature. However, some carbon abundance derivations in such dwarfs have been reported \citep[e.g.,][]{Suarez-17} since the pioneering works of \citet{Pagel-64} and \citet{Cohen-86}. Some measurements of carbon isotopic ratios are reported in giant stars with lower $^{12}$C/$^{13}$C ratios, and thus stronger $^{13}$CH lines \citep[see, e.g.,][]{Sneden-86, Aoki-06}.

Our analysis has been conducted by computing synthetic spectra with the TURBO spectrum code \citep{Alvarez-98} and  MARCS model atmospheres \citep{Gustafsson-08} interpolated at the stellar parameters of the sibling candidates. We adopted the CH linelist (including $^{12}$CH and $^{13}$CH lines) from  \citet{Masseron-14} and the Vienna Atomic Line Database\footnote{VALD: http://vald.inasan.ru/~vald3/php/vald.php} \citep[VALD;][]{Piskunov-95, Kupka-99, Ryabchikova-15}. First, we estimated the carbon abundance by comparing the synthetic and observed spectra over a wide wavelength domain around the G band (typically from 4290 to 4320 \AA). After fixing the carbon abundance, we examined several spectral regions where the $^{13}$CH lines are predicted to be stronger. We point out that since the C$^{13}$H lines become extremely weak for $^{12}$C/$^{13}$C ratios higher than $\sim$60, only a lower limit can be estimated. To validate our procedure, we first analyzed a HARPS spectrum of Vesta and derived A(C) = 8.35 and a lower limit of $^{12}$C/$^{13}$C $\gtrsim$ 70 for the Sun, in agreement with the reference solar values: A(C) = 8.39~dex \citep{Asplund-05} and $^{12}$C/$^{13}$C = 86.8 \citep{Scott-06}. This method allowed us to derive a carbon abundance equal to 8.40 and 8.35~dex for HD186302 and HD52456, respectively (i.e., a carbon abundance very close to that of the Sun, which confirs their probable sibling nature). Furthermore, their carbon isotopic ratio is found to be higher than $\sim$50-60. As for the solar spectrum, their $^{13}$CH lines are too weak to analyze them well in crowded spectra, which are difficult to normalize. Moreover, because of such difficulties, it was impossible to derive abundances for the spectra of our two other sibling candidates since they are even cooler and thus suffer from much stronger blends and normalization problems. The reported carbon abundances and upper limits for the $^{12}$C/$^{13}$C ratio in HD186302 and HD52456 agree with the hypothesis that these stars are solar siblings.

\section{Summary} \label{discussion}

As discussed in the introduction, finding solar siblings will help us to better understand the environment in which our Sun has been formed. The search for solar siblings is difficult since these stars are expected to have been spread throughout the Milky Way. Only a handful of solar siblings are expected (depending on how they are dispersed in the Galaxy) to be found in the solar vicinity \citep{Zwart-09}. Moreover, most of these siblings are expected to have low masses (because of the initial mass function) and thus be faint \citep{Martinez-Barbosa-16}, which means that they are hard to find and then to precisely characterize. Since by definition solar siblings are the stars that were formed with the Sun in the same cluster, i) their chemical abundances must be very similar to those of the Sun (if stellar evolution did not affect the atmospheric chemical composition)  and ii) their age must be the same as that of the Sun. In addition, they should have present-day kinematics, suggesting that they have been formed in the same location as the Sun. The predictions about the kinematics and orbits of solar siblings in the vicinity of our Sun is difficult and heavily depends on the characteristics of the solar birth cluster (size, location, etc.) and the adopted Galactic model. We here carried out a mostly chemical and age-based search for solar siblings among the AMBRE stars that are located within about 200 pc ($\varpi$ $\geq$ 5mas). The selected sibling candidates were then briefly studied to understand how similar are they to the Sun in terms of kinematics.

From about 17,000 stars characterized within the AMBRE program, we selected 55 stars with a stellar metallicity ([Fe/H]) from $-$0.1 to 0.1 dex. For these stars we derived the chemical abundances of several iron-peak, $\alpha$-capture, and neutron-capture elements. Then we selected 12 stars whose average abundances and metallicity are very similar to those of the Sun (by $\pm$0.03 dex).

The stellar ages of these 12 sibling candidates, derived by chemistry, allowed us to narrow the 12 stars down to a sample of 4 candidates whose ages are similar to that of the Sun within the uncertainties. The ages of the stars were derived by isochrone fitting. Additional independent age indicators (such as chromospheric activity, Li abundance, and age-sensitive abundance ratios) were also used to confirm or refute the isochronal ages of the stars with large uncertainties. Of the 4 selected stars, HD186302 alone has an isochronal age error of less than  1 Gyr. The other 3 candidates have errors in age greater than 3 Gyr.

Our very simple kinematic study of the four best-fit candidates suggests that HD176535 might be the least likely candidate. HD186302 is the most likely candidate in terms of kinematics.

As a final test, we derived the carbon isotopic ratio for the two hottest (\teff \ $>$ 5000 K) candidates that were compatible with the solar value. We suggest that other isotopic ratios sensitive to Galactic chemical evolution such as $^{24}$Mg/$^{25}$Mg and $^{25}$Mg/$^{26}$Mg \citep[][]{Kobayashi-11} can be used to identify solar siblings. For example, \citet{Yong-04} have studied the $^{24}$Mg/$^{25}$Mg and $^{25}$Mg/$^{26}$Mg ratios in Hyades cool dwarfs and found a good agreement in the values of isotope ratios in the stars belonging to the cluster.

We note that we did not consider  some important astrophysical effects such as atomic diffusion and extra mixing mechanisms that can affect the atmospheric abundances of the stars during their evolution and also influence the derivation of their isochronal ages \citep[e.g.,][]{Dotter-17}. We did not consider either the possibility of a change in stellar surface abundances caused by planet engulfment and/or planet formation \citep{Smith-01, Melendez-09, Ramirez-10}. Some theoretical works support the possible modification of stellar abundances through planet formation \citep[e.g.,][]{Kunitomo-18, Chambers-10}, \citep[but see also][]{Theado-12, Gustafsson-18}. However, from an observational point of view, the presence of chemical signatures of planet formation in the spectra of stars is still lively debated \citep[see, e.g.,][and references therein]{Adibekyan-14, Adibekyan-17}.

In summary, we identified four solar sibling candidates in the AMBRE data. Because of its relatively high temperature (\teff \ = 5675 K), HD186302 is the most precisely characterized star of the four. Interestingly, HD176535 has previously been identified as a solar sibling candidate by \citet{Batista-14}. The other three candidates are cooler, which led to larger errors in the atmospheric parameters, chemical abundances, and ages. Additional spectroscopic characterization of these stars (e.g., using the ESPRESSO ultrahigh-resolution spectra) including derivation of different isotopic ratios will help to confirm or refute their relation with the solar parent cluster.

\begin{acknowledgements}
{This work was supported by Funda\c{c}\~ao para a Ci\^encia e Tecnologia (FCT) through national funds (project ref. PTDC/FIS-AST/7073/2014, PTDC/FIS-AST/32113/2017, and PTDC/FIS-AST/32113/2017)   and by Fundo Europeu de Desenvolvimento Regional (FEDER) through the COMPETE 2020 - Programa Operacional Competitividade e Internacionalização (POCI) (project ref. POCI-01-0145-FEDER-007672, POCI-01-0145-FEDER-032113). This work was also supported by FCT, POCI, FEDER in the framework of the project POCI-01-0145-FEDER-032113.  V.A., E.D.M, N.C.S., and S.G.S. also acknowledge the support from FCT through Investigador FCT contracts nr.  IF/00650/2015/CP1273/CT0001, IF/00849/2015/CP1273/CT0003, IF/00169/2012/CP0150/CT0002,   and IF/00028/2014/CP1215/CT0002, respectively, and POPH/FSE (EC) by FEDER funding through the program ``Programa Operacional de Factores de Competitividade - COMPETE''.  V.A. also acknowledges the support from Observatoire de la C\^ote d'Azur (OCA) for the visitor program and would like to offer his special thanks to the OCA stuff for all the support during his stay in Nice.  P.deL., A.R.B, and G.K. acknowledge financial support form the ANR 14-CE33-014-01. ACSF is supported by grant 234989/2014-9 from CNPq (Brazil).}
\end{acknowledgements}
  %________________________________________________________________

\bibliography{references}

\begin{thebibliography}{128}
\expandafter\ifx\csname natexlab\endcsname\relax\def\natexlab#1{#1}\fi

\bibitem[{{Adams}(2010)}]{Adams-10}
{Adams}, F.~C. 2010, \araa, 48, 47

\bibitem[{{Adams} \& {Spergel}(2005)}]{Adams-05}
{Adams}, F.~C. \& {Spergel}, D.~N. 2005, Astrobiology, 5, 497

\bibitem[{{Adibekyan} {et~al.}(2017){Adibekyan}, {Delgado-Mena}, {Feltzing},
  {Gonz{\'a}lez Hern{\'a}ndez}, {Hinkel}, {Korn}, {Asplund}, {Beck}, {Deal},
  {Gustafsson}, {Honda}, {Lind}, {Nissen}, \& {Spina}}]{Adibekyan-17}
{Adibekyan}, V., {Delgado-Mena}, E., {Feltzing}, S., {et~al.} 2017,
  Astronomische Nachrichten, 338, 442

\bibitem[{{Adibekyan} {et~al.}(2016){Adibekyan}, {Delgado-Mena}, {Figueira},
  {Sousa}, {Santos}, {Faria}, {Gonz{\'a}lez Hern{\'a}ndez}, {Israelian},
  {Harutyunyan}, {Su{\'a}rez-Andr{\'e}s}, \& {Hakobyan}}]{Adibekyan-16}
{Adibekyan}, V., {Delgado-Mena}, E., {Figueira}, P., {et~al.} 2016, \aap, 591,
  A34

\bibitem[{{Adibekyan} {et~al.}(2015{\natexlab{a}}){Adibekyan}, {Figueira},
  {Santos}, {Sousa}, {Faria}, {Delgado-Mena}, {Oshagh}, {Tsantaki}, {Hakobyan},
  {Gonz{\'a}lez Hern{\'a}ndez}, {Su{\'a}rez-Andr{\'e}s}, \&
  {Israelian}}]{Adibekyan-15}
{Adibekyan}, V., {Figueira}, P., {Santos}, N.~C., {et~al.} 2015{\natexlab{a}},
  \aap, 583, A94

\bibitem[{{Adibekyan} {et~al.}(2015{\natexlab{b}}){Adibekyan}, {Benamati},
  {Santos}, {Alves}, {Lovis}, {Udry}, {Israelian}, {Sousa}, {Tsantaki},
  {Mortier}, {Sozzetti}, \& {De Medeiros}}]{Adibekyan-15a}
{Adibekyan}, V.~Z., {Benamati}, L., {Santos}, N.~C., {et~al.}
  2015{\natexlab{b}}, \mnras, 450, 1900

\bibitem[{{Adibekyan} {et~al.}(2014){Adibekyan}, {Gonz{\'a}lez Hern{\'a}ndez},
  {Delgado Mena}, {Sousa}, {Santos}, {Israelian}, {Figueira}, \& {Bertran de
  Lis}}]{Adibekyan-14}
{Adibekyan}, V.~Z., {Gonz{\'a}lez Hern{\'a}ndez}, J.~I., {Delgado Mena}, E.,
  {et~al.} 2014, \aap, 564, L15

\bibitem[{{Adibekyan} {et~al.}(2012){Adibekyan}, {Sousa}, {Santos}, {Delgado
  Mena}, {Gonz{\'a}lez Hern{\'a}ndez}, {Israelian}, {Mayor}, \&
  {Khachatryan}}]{Adibekyan-12}
{Adibekyan}, V.~Z., {Sousa}, S.~G., {Santos}, N.~C., {et~al.} 2012, \aap, 545,
  A32

\bibitem[{{Alexeeva} {et~al.}(2014){Alexeeva}, {Pakhomov}, \&
  {Mashonkina}}]{Alexeeva-14}
{Alexeeva}, S.~A., {Pakhomov}, Y.~V., \& {Mashonkina}, L.~I. 2014, Astronomy
  Letters, 40, 406

\bibitem[{{Alvarez} \& {Plez}(1998)}]{Alvarez-98}
{Alvarez}, R. \& {Plez}, B. 1998, \aap, 330, 1109

\bibitem[{{Andr{\'a}ssy} \& {Spruit}(2015)}]{Andrassy-15}
{Andr{\'a}ssy}, R. \& {Spruit}, H.~C. 2015, \aap, 579, A122

\bibitem[{{Aoki} {et~al.}(2006){Aoki}, {Frebel}, {Christlieb}, {Norris},
  {Beers}, {Minezaki}, {Barklem}, {Honda}, {Takada-Hidai}, {Asplund}, {Ryan},
  {Tsangarides}, {Eriksson}, {Steinhauer}, {Deliyannis}, {Nomoto}, {Fujimoto},
  {Ando}, {Yoshii}, \& {Kajino}}]{Aoki-06}
{Aoki}, W., {Frebel}, A., {Christlieb}, N., {et~al.} 2006, \apj, 639, 897

\bibitem[{{Arenou} {et~al.}(2018){Arenou}, {Luri}, {Babusiaux}, {Fabricius},
  {Helmi}, {Muraveva}, {Robin}, {Spoto}, {Vallenari}, {Antoja},
  {Cantat-Gaudin}, {Jordi}, {Leclerc}, {Reyl{\'e}}, {Romero-G{\'o}mez}, {Shih},
  {Soria}, {Barache}, {Bossini}, {Bragaglia}, {Breddels}, {Fabrizio},
  {Lambert}, {Marrese}, {Massari}, {Moitinho}, {Robichon}, {Ruiz-Dern},
  {Sordo}, {Veljanoski}, {Di Matteo}, {Eyer}, {Jasniewicz}, {Pancino},
  {Soubiran}, {Spagna}, {Tanga}, {Turon}, \& {Zurbach}}]{Arenou-18}
{Arenou}, F., {Luri}, X., {Babusiaux}, C., {et~al.} 2018, ArXiv e-prints
  [\eprint[arXiv]{1804.09375}]

\bibitem[{{Armillotta} {et~al.}(2018){Armillotta}, {Krumholz}, \&
  {Fujimoto}}]{Armillotta-18}
{Armillotta}, L., {Krumholz}, M.~R., \& {Fujimoto}, Y. 2018, ArXiv e-prints
  [\eprint[arXiv]{1807.01712}]

\bibitem[{{Asplund} {et~al.}(2005){Asplund}, {Grevesse}, {Sauval}, {Allende
  Prieto}, \& {Blomme}}]{Asplund-05}
{Asplund}, M., {Grevesse}, N., {Sauval}, A.~J., {Allende Prieto}, C., \&
  {Blomme}, R. 2005, \aap, 431, 693

\bibitem[{{Batista} {et~al.}(2014){Batista}, {Adibekyan}, {Sousa}, {Santos},
  {Delgado Mena}, \& {Hakobyan}}]{Batista-14}
{Batista}, S.~F.~A., {Adibekyan}, V.~Z., {Sousa}, S.~G., {et~al.} 2014, \aap,
  564, A43

\bibitem[{{Batista} \& {Fernandes}(2012)}]{Batista-12}
{Batista}, S.~F.~A. \& {Fernandes}, J. 2012, \na, 17, 514

\bibitem[{{Bedell} {et~al.}(2014){Bedell}, {Mel{\'e}ndez}, {Bean},
  {Ram{\'{\i}}rez}, {Leite}, \& {Asplund}}]{Bedell-14}
{Bedell}, M., {Mel{\'e}ndez}, J., {Bean}, J.~L., {et~al.} 2014, \apj, 795, 23

\bibitem[{{Bensby} {et~al.}(2014){Bensby}, {Feltzing}, \& {Oey}}]{Bensby-14}
{Bensby}, T., {Feltzing}, S., \& {Oey}, M.~S. 2014, \aap, 562, A71

\bibitem[{{Bertran de Lis} {et~al.}(2015){Bertran de Lis}, {Delgado Mena},
  {Adibekyan}, {Santos}, \& {Sousa}}]{Bertran-15}
{Bertran de Lis}, S., {Delgado Mena}, E., {Adibekyan}, V.~Z., {Santos}, N.~C.,
  \& {Sousa}, S.~G. 2015, \aap, 576, A89

\bibitem[{{Bland-Hawthorn} {et~al.}(2010){Bland-Hawthorn}, {Krumholz}, \&
  {Freeman}}]{Bland-Hawthorn-10}
{Bland-Hawthorn}, J., {Krumholz}, M.~R., \& {Freeman}, K. 2010, \apj, 713, 166

\bibitem[{{Bobylev} \& {Bajkova}(2014)}]{Bobylev-14}
{Bobylev}, V.~V. \& {Bajkova}, A.~T. 2014, Astronomy Letters, 40, 353

\bibitem[{{Bobylev} {et~al.}(2011){Bobylev}, {Bajkova}, {Myll{\"a}ri}, \&
  {Valtonen}}]{Bobylev-11}
{Bobylev}, V.~V., {Bajkova}, A.~T., {Myll{\"a}ri}, A., \& {Valtonen}, M. 2011,
  Astronomy Letters, 37, 550

\bibitem[{{Bonanno} \& {Fr{\"o}hlich}(2015)}]{Bonanno-15}
{Bonanno}, A. \& {Fr{\"o}hlich}, H.-E. 2015, \aap, 580, A130

\bibitem[{{Bovy}(2015)}]{Bovy-15}
{Bovy}, J. 2015, \apjs, 216, 29

\bibitem[{{Bovy}(2016)}]{Bovy-16}
{Bovy}, J. 2016, \apj, 817, 49

\bibitem[{{Bressan} {et~al.}(2012){Bressan}, {Marigo}, {Girardi}, {Salasnich},
  {Dal Cero}, {Rubele}, \& {Nanni}}]{Bressan-12}
{Bressan}, A., {Marigo}, P., {Girardi}, L., {et~al.} 2012, \mnras, 427, 127

\bibitem[{{Brown} {et~al.}(2010){Brown}, {Portegies Zwart}, \&
  {Bean}}]{Brown-10}
{Brown}, A.~G.~A., {Portegies Zwart}, S.~F., \& {Bean}, J. 2010, \mnras, 407,
  458

\bibitem[{{Brown} {et~al.}(2004){Brown}, {Trujillo}, \&
  {Rabinowitz}}]{Brown-04}
{Brown}, M.~E., {Trujillo}, C., \& {Rabinowitz}, D. 2004, \apj, 617, 645

\bibitem[{{Cayrel}(1988)}]{Cayrel-88}
{Cayrel}, R. 1988, in IAU Symposium, Vol. 132, The Impact of Very High S/N
  Spectroscopy on Stellar Physics, ed. G.~{Cayrel de Strobel} \& M.~{Spite},
  345

\bibitem[{{Chabrier}(2001)}]{Chabrier-01}
{Chabrier}, G. 2001, \apj, 554, 1274

\bibitem[{{Chambers}(2010)}]{Chambers-10}
{Chambers}, J.~E. 2010, \apj, 724, 92

\bibitem[{{Charbonnel} \& {Talon}(2005)}]{Charbonnel-05}
{Charbonnel}, C. \& {Talon}, S. 2005, Science, 309, 2189

\bibitem[{{Cohen}(1968)}]{Cohen-86}
{Cohen}, J.~G. 1968, \aplett, 2, 163

\bibitem[{{Cosentino} {et~al.}(2012){Cosentino}, {Lovis}, {Pepe}, {Collier
  Cameron}, {Latham}, {Molinari}, {Udry}, {Bezawada}, {Black}, {Born},
  {Buchschacher}, {Charbonneau}, {Figueira}, {Fleury}, {Galli}, {Gallie},
  {Gao}, {Ghedina}, {Gonzalez}, {Gonzalez}, {Guerra}, {Henry}, {Horne},
  {Hughes}, {Kelly}, {Lodi}, {Lunney}, {Maire}, {Mayor}, {Micela}, {Ordway},
  {Peacock}, {Phillips}, {Piotto}, {Pollacco}, {Queloz}, {Rice}, {Riverol},
  {Riverol}, {San Juan}, {Sasselov}, {Segransan}, {Sozzetti}, {Sosnowska},
  {Stobie}, {Szentgyorgyi}, {Vick}, \& {Weber}}]{Cosentino-12}
{Cosentino}, R., {Lovis}, C., {Pepe}, F., {et~al.} 2012, in \procspie, Vol.
  8446, Ground-based and Airborne Instrumentation for Astronomy IV, 84461V

\bibitem[{{da Silva} {et~al.}(2006){da Silva}, {Girardi}, {Pasquini},
  {Setiawan}, {von der L{\"u}he}, {de Medeiros}, {Hatzes}, {D{\"o}llinger}, \&
  {Weiss}}]{daSilva-06}
{da Silva}, L., {Girardi}, L., {Pasquini}, L., {et~al.} 2006, \aap, 458, 609

\bibitem[{{de Laverny} {et~al.}(2013){de Laverny}, {Recio-Blanco}, {Worley},
  {De Pascale}, {Hill}, \& {Bijaoui}}]{deLaverny-13}
{de Laverny}, P., {Recio-Blanco}, A., {Worley}, C.~C., {et~al.} 2013, The
  Messenger, 153, 18

\bibitem[{{de Laverny} {et~al.}(2012){de Laverny}, {Recio-Blanco}, {Worley}, \&
  {Plez}}]{deLaverny-12}
{de Laverny}, P., {Recio-Blanco}, A., {Worley}, C.~C., \& {Plez}, B. 2012,
  \aap, 544, A126

\bibitem[{{De Pascale} {et~al.}(2014){De Pascale}, {Worley}, {de Laverny},
  {Recio-Blanco}, {Hill}, \& {Bijaoui}}]{DePascale-14}
{De Pascale}, M., {Worley}, C.~C., {de Laverny}, P., {et~al.} 2014, \aap, 570,
  A68

\bibitem[{{Delgado Mena} {et~al.}(2015){Delgado Mena}, {Bertr{\'a}n de Lis},
  {Adibekyan}, {Sousa}, {Figueira}, {Mortier}, {Gonz{\'a}lez Hern{\'a}ndez},
  {Tsantaki}, {Israelian}, \& {Santos}}]{Delgado-Mena-15}
{Delgado Mena}, E., {Bertr{\'a}n de Lis}, S., {Adibekyan}, V.~Z., {et~al.}
  2015, \aap, 576, A69

\bibitem[{{Delgado Mena} {et~al.}(2014){Delgado Mena}, {Israelian},
  {Gonz{\'a}lez Hern{\'a}ndez}, {Sousa}, {Mortier}, {Santos}, {Adibekyan},
  {Fernandes}, {Rebolo}, {Udry}, \& {Mayor}}]{Delgado-Mena-14}
{Delgado Mena}, E., {Israelian}, G., {Gonz{\'a}lez Hern{\'a}ndez}, J.~I.,
  {et~al.} 2014, \aap, 562, A92

\bibitem[{{Delgado Mena} {et~al.}(2017){Delgado Mena}, {Tsantaki}, {Adibekyan},
  {Sousa}, {Santos}, {Gonz{\'a}lez Hern{\'a}ndez}, \&
  {Israelian}}]{Delgado-Mena-17}
{Delgado Mena}, E., {Tsantaki}, M., {Adibekyan}, V.~Z., {et~al.} 2017, \aap,
  606, A94

\bibitem[{{Delgado Mena} {et~al.}(2018){Delgado Mena}, {Tsantaki},
  {Zh.~Adibekyan}, {Sousa}, {Santos}, {Gonz{\'a}lez Hern{\'a}ndez}, \&
  {Israelian}}]{Delgado-Mena-18}
{Delgado Mena}, E., {Tsantaki}, M., {Zh.~Adibekyan}, V., {et~al.} 2018, in IAU
  Symposium, Vol. 330, IAU Symposium, ed. A.~{Recio-Blanco}, P.~{de Laverny},
  A.~G.~A. {Brown}, \& T.~{Prusti}, 156--159

\bibitem[{{Dotter} {et~al.}(2017){Dotter}, {Conroy}, {Cargile}, \&
  {Asplund}}]{Dotter-17}
{Dotter}, A., {Conroy}, C., {Cargile}, P., \& {Asplund}, M. 2017, \apj, 840, 99

\bibitem[{{Feng} \& {Krumholz}(2014)}]{Feng-14}
{Feng}, Y. \& {Krumholz}, M.~R. 2014, \nat, 513, 523

\bibitem[{{Flower}(1996)}]{Flower-96}
{Flower}, P.~J. 1996, \apj, 469, 355

\bibitem[{{Gaia Collaboration} {et~al.}(2018){Gaia Collaboration}, {Brown},
  {Vallenari}, {Prusti}, {de Bruijne}, {Babusiaux}, \&
  {Bailer-Jones}}]{Gaia-18}
{Gaia Collaboration}, {Brown}, A.~G.~A., {Vallenari}, A., {et~al.} 2018, ArXiv
  e-prints [\eprint[arXiv]{1804.09365}]

\bibitem[{{Garcia-Lopez} {et~al.}(1993){Garcia-Lopez}, {Rebolo}, {Beckman}, \&
  {McKeith}}]{Garcia-Lopez-93}
{Garcia-Lopez}, R.~J., {Rebolo}, R., {Beckman}, J.~E., \& {McKeith}, C.~D.
  1993, \aap, 273, 482

\bibitem[{{Gustafsson}(2018)}]{Gustafsson-18}
{Gustafsson}, B. 2018, ArXiv e-prints [\eprint[arXiv]{1809.02361}]

\bibitem[{{Gustafsson} {et~al.}(2008){Gustafsson}, {Edvardsson}, {Eriksson},
  {J{\o}rgensen}, {Nordlund}, \& {Plez}}]{Gustafsson-08}
{Gustafsson}, B., {Edvardsson}, B., {Eriksson}, K., {et~al.} 2008, \aap, 486,
  951

\bibitem[{{Hogg} {et~al.}(2016){Hogg}, {Casey}, {Ness}, {Rix},
  {Foreman-Mackey}, {Hasselquist}, {Ho}, {Holtzman}, {Majewski}, {Martell},
  {M{\'e}sz{\'a}ros}, {Nidever}, \& {Shetrone}}]{Hogg-16}
{Hogg}, D.~W., {Casey}, A.~R., {Ness}, M., {et~al.} 2016, \apj, 833, 262

\bibitem[{{Jacobson} {et~al.}(2007){Jacobson}, {Friel}, \&
  {Pilachowski}}]{Jacobson-07}
{Jacobson}, H.~R., {Friel}, E.~D., \& {Pilachowski}, C.~A. 2007, \aj, 134, 1216

\bibitem[{{Kobayashi} {et~al.}(2011){Kobayashi}, {Karakas}, \&
  {Umeda}}]{Kobayashi-11}
{Kobayashi}, C., {Karakas}, A.~I., \& {Umeda}, H. 2011, \mnras, 414, 3231

\bibitem[{{Kordopatis} {et~al.}(2011){Kordopatis}, {Recio-Blanco}, {de
  Laverny}, {Bijaoui}, {Hill}, {Gilmore}, {Wyse}, \&
  {Ordenovic}}]{Kordopatis-11}
{Kordopatis}, G., {Recio-Blanco}, A., {de Laverny}, P., {et~al.} 2011, \aap,
  535, A106

\bibitem[{{Kunitomo} {et~al.}(2018){Kunitomo}, {Guillot}, {Ida}, \&
  {Takeuchi}}]{Kunitomo-18}
{Kunitomo}, M., {Guillot}, T., {Ida}, S., \& {Takeuchi}, T. 2018, ArXiv
  e-prints [\eprint[arXiv]{1808.07396}]

\bibitem[{{Kupka} {et~al.}(1999){Kupka}, {Piskunov}, {Ryabchikova}, {Stempels},
  \& {Weiss}}]{Kupka-99}
{Kupka}, F., {Piskunov}, N., {Ryabchikova}, T.~A., {Stempels}, H.~C., \&
  {Weiss}, W.~W. 1999, \aaps, 138, 119

\bibitem[{{Kurucz}(1993)}]{Kurucz-93}
{Kurucz}, R.~L. 1993, {SYNTHE spectrum synthesis programs and line data}

\bibitem[{{Lada} \& {Lada}(2003)}]{Lada-03}
{Lada}, C.~J. \& {Lada}, E.~A. 2003, \araa, 41, 57

\bibitem[{{Lamers} \& {Gieles}(2006)}]{Lamers-06}
{Lamers}, H.~J.~G.~L.~M. \& {Gieles}, M. 2006, \aap, 455, L17

\bibitem[{{Lindegren} {et~al.}(2018){Lindegren}, {Hernandez}, {Bombrun},
  {Klioner}, {Bastian}, {Ramos-Lerate}, {de Torres}, {Steidelmuller},
  {Stephenson}, {Hobbs}, {Lammers}, {Biermann}, {Geyer}, {Hilger}, {Michalik},
  {Stampa}, {McMillan}, {Castaneda}, {Clotet}, {Comoretto}, {Davidson},
  {Fabricius}, {Gracia}, {Hambly}, {Hutton}, {Mora}, {Portell}, {van Leeuwen},
  {Abbas}, {Abreu}, {Altmann}, {Andrei}, {Anglada}, {Balaguer-Nunez},
  {Barache}, {Becciani}, {Bertone}, {Bianchi}, {Bouquillon}, {Bourda},
  {Brusemeister}, {Bucciarelli}, {Busonero}, {Buzzi}, {Cancelliere},
  {Carlucci}, {Charlot}, {Cheek}, {Crosta}, {Crowley}, {de Bruijne}, {de
  Felice}, {Drimmel}, {Esquej}, {Fienga}, {Fraile}, {Gai}, {Garralda},
  {Gonzalez-Vidal}, {Guerra}, {Hauser}, {Hofmann}, {Holl}, {Jordan},
  {Lattanzi}, {Lenhardt}, {Liao}, {Licata}, {Lister}, {Loffler}, {Marchant},
  {Martin-Fleitas}, {Messineo}, {Mignard}, {Morbidelli}, {Poggio}, {Riva},
  {Rowell}, {Salguero}, {Sarasso}, {Sciacca}, {Siddiqui}, {Smart}, {Spagna},
  {Steele}, {Taris}, {Torra}, {van Elteren}, {van Reeven}, \&
  {Vecchiato}}]{Lindegren-18}
{Lindegren}, L., {Hernandez}, J., {Bombrun}, A., {et~al.} 2018, ArXiv e-prints
  [\eprint[arXiv]{1804.09366}]

\bibitem[{{Liu} {et~al.}(2015){Liu}, {Ruchti}, {Feltzing},
  {Mart{\'{\i}}nez-Barbosa}, {Bensby}, {Brown}, \& {Portegies Zwart}}]{Liu-15}
{Liu}, C., {Ruchti}, G., {Feltzing}, S., {et~al.} 2015, \aap, 575, A51

\bibitem[{{Liu} {et~al.}(2016{\natexlab{a}}){Liu}, {Asplund}, {Yong},
  {Mel{\'e}ndez}, {Ram{\'{\i}}rez}, {Karakas}, {Carlos}, \& {Marino}}]{Liu-16a}
{Liu}, F., {Asplund}, M., {Yong}, D., {et~al.} 2016{\natexlab{a}}, \mnras, 463,
  696

\bibitem[{{Liu} {et~al.}(2016{\natexlab{b}}){Liu}, {Yong}, {Asplund},
  {Ram{\'{\i}}rez}, \& {Mel{\'e}ndez}}]{Liu-16}
{Liu}, F., {Yong}, D., {Asplund}, M., {Ram{\'{\i}}rez}, I., \& {Mel{\'e}ndez},
  J. 2016{\natexlab{b}}, \mnras, 457, 3934

\bibitem[{{Looney} {et~al.}(2006){Looney}, {Tobin}, \& {Fields}}]{Looney-06}
{Looney}, L.~W., {Tobin}, J.~J., \& {Fields}, B.~D. 2006, \apj, 652, 1755

\bibitem[{{Lorenzo-Oliveira} {et~al.}(2018){Lorenzo-Oliveira}, {Freitas},
  {Mel{\'e}ndez}, {Bedell}, {Ram{\'{\i}}rez}, {Bean}, {Asplund}, {Spina},
  {Dreizler}, {Alves-Brito}, \& {Casagrande}}]{Lorenzo-Oliveira-18}
{Lorenzo-Oliveira}, D., {Freitas}, F.~C., {Mel{\'e}ndez}, J., {et~al.} 2018,
  ArXiv e-prints [\eprint[arXiv]{1806.08014}]

\bibitem[{{Lorenzo-Oliveira} {et~al.}(2016){Lorenzo-Oliveira}, {Porto de
  Mello}, \& {Schiavon}}]{Lorenzo-Oliveira-16}
{Lorenzo-Oliveira}, D., {Porto de Mello}, G.~F., \& {Schiavon}, R.~P. 2016,
  \aap, 594, L3

\bibitem[{{Luri} {et~al.}(2018){Luri}, {Brown}, {Sarro}, {Arenou},
  {Bailer-Jones}, {Castro-Ginard}, {de Bruijne}, {Prusti}, {Babusiaux}, \&
  {Delgado}}]{Luri-18}
{Luri}, X., {Brown}, A.~G.~A., {Sarro}, L.~M., {et~al.} 2018, \aap, 616, A9

\bibitem[{{Mamajek} \& {Hillenbrand}(2008)}]{Mamajek-08}
{Mamajek}, E.~E. \& {Hillenbrand}, L.~A. 2008, \apj, 687, 1264

\bibitem[{{Mart{\'{\i}}nez-Barbosa} {et~al.}(2016){Mart{\'{\i}}nez-Barbosa},
  {Brown}, {Boekholt}, {Portegies Zwart}, {Antiche}, \&
  {Antoja}}]{Martinez-Barbosa-16}
{Mart{\'{\i}}nez-Barbosa}, C.~A., {Brown}, A.~G.~A., {Boekholt}, T., {et~al.}
  2016, \mnras, 457, 1062

\bibitem[{{Masseron} {et~al.}(2014){Masseron}, {Plez}, {Van Eck}, {Colin},
  {Daoutidis}, {Godefroid}, {Coheur}, {Bernath}, {Jorissen}, \&
  {Christlieb}}]{Masseron-14}
{Masseron}, T., {Plez}, B., {Van Eck}, S., {et~al.} 2014, \aap, 571, A47

\bibitem[{{Mel{\'e}ndez} {et~al.}(2009){Mel{\'e}ndez}, {Asplund}, {Gustafsson},
  \& {Yong}}]{Melendez-09}
{Mel{\'e}ndez}, J., {Asplund}, M., {Gustafsson}, B., \& {Yong}, D. 2009, \apjl,
  704, L66

\bibitem[{{Mel{\'e}ndez} {et~al.}(2014){Mel{\'e}ndez}, {Ram{\'{\i}}rez},
  {Karakas}, {Yong}, {Monroe}, {Bedell}, {Bergemann}, {Asplund}, {Tucci Maia},
  {Bean}, {do Nascimento}, {Bazot}, {Alves-Brito}, {Freitas}, \&
  {Castro}}]{Melendez-14}
{Mel{\'e}ndez}, J., {Ram{\'{\i}}rez}, I., {Karakas}, A.~I., {et~al.} 2014,
  \apj, 791, 14

\bibitem[{{Mikolaitis} {et~al.}(2017){Mikolaitis}, {de Laverny},
  {Recio-Blanco}, {Hill}, {Worley}, \& {de Pascale}}]{Mikolaitis-17}
{Mikolaitis}, {\v S}., {de Laverny}, P., {Recio-Blanco}, A., {et~al.} 2017,
  \aap, 600, A22

\bibitem[{{Morbidelli} \& {Levison}(2004)}]{Morbidelli-04}
{Morbidelli}, A. \& {Levison}, H.~F. 2004, \aj, 128, 2564

\bibitem[{{Mortier} {et~al.}(2013){Mortier}, {Santos}, {Sousa}, {Fernandes},
  {Adibekyan}, {Delgado Mena}, {Montalto}, \& {Israelian}}]{Mortier-13}
{Mortier}, A., {Santos}, N.~C., {Sousa}, S.~G., {et~al.} 2013, \aap, 558, A106

\bibitem[{{Mortier} {et~al.}(2014){Mortier}, {Sousa}, {Adibekyan},
  {Brand{\~a}o}, \& {Santos}}]{Mortier-14}
{Mortier}, A., {Sousa}, S.~G., {Adibekyan}, V.~Z., {Brand{\~a}o}, I.~M., \&
  {Santos}, N.~C. 2014, \aap, 572, A95

\bibitem[{{Ness} {et~al.}(2018){Ness}, {Rix}, {Hogg}, {Casey}, {Holtzman},
  {Fouesneau}, {Zasowski}, {Geisler}, {Shetrone}, {Minniti}, {Frinchaboy}, \&
  {Roman-Lopes}}]{Ness-18}
{Ness}, M., {Rix}, H.-W., {Hogg}, D.~W., {et~al.} 2018, \apj, 853, 198

\bibitem[{{Nissen}(2015)}]{Nissen-15}
{Nissen}, P.~E. 2015, \aap, 579, A52

\bibitem[{{Noyes} {et~al.}(1984){Noyes}, {Hartmann}, {Baliunas}, {Duncan}, \&
  {Vaughan}}]{Noyes-84}
{Noyes}, R.~W., {Hartmann}, L.~W., {Baliunas}, S.~L., {Duncan}, D.~K., \&
  {Vaughan}, A.~H. 1984, \apj, 279, 763

\bibitem[{{Pace}(2013)}]{Pace-13}
{Pace}, G. 2013, \aap, 551, L8

\bibitem[{{Pagel}(1964)}]{Pagel-64}
{Pagel}, B.~E.~J. 1964, Royal Greenwich Observatory Bulletins, 87, 227

\bibitem[{{Piskunov} {et~al.}(2006){Piskunov}, {Kharchenko}, {R{\"o}ser},
  {Schilbach}, \& {Scholz}}]{Piskunov-06}
{Piskunov}, A.~E., {Kharchenko}, N.~V., {R{\"o}ser}, S., {Schilbach}, E., \&
  {Scholz}, R.-D. 2006, \aap, 445, 545

\bibitem[{{Piskunov} {et~al.}(1995){Piskunov}, {Kupka}, {Ryabchikova}, {Weiss},
  \& {Jeffery}}]{Piskunov-95}
{Piskunov}, N.~E., {Kupka}, F., {Ryabchikova}, T.~A., {Weiss}, W.~W., \&
  {Jeffery}, C.~S. 1995, \aaps, 112, 525

\bibitem[{{Portegies Zwart}(2009)}]{Zwart-09}
{Portegies Zwart}, S.~F. 2009, \apjl, 696, L13

\bibitem[{{Ram{\'{\i}}rez} {et~al.}(2010){Ram{\'{\i}}rez}, {Asplund},
  {Baumann}, {Mel{\'e}ndez}, \& {Bensby}}]{Ramirez-10}
{Ram{\'{\i}}rez}, I., {Asplund}, M., {Baumann}, P., {Mel{\'e}ndez}, J., \&
  {Bensby}, T. 2010, \aap, 521, A33

\bibitem[{{Ram{\'{\i}}rez} {et~al.}(2014){Ram{\'{\i}}rez}, {Bajkova},
  {Bobylev}, {Roederer}, {Lambert}, {Endl}, {Cochran}, {MacQueen}, \&
  {Wittenmyer}}]{Ramirez-14}
{Ram{\'{\i}}rez}, I., {Bajkova}, A.~T., {Bobylev}, V.~V., {et~al.} 2014, \apj,
  787, 154

\bibitem[{{Recio-Blanco} {et~al.}(2006){Recio-Blanco}, {Bijaoui}, \& {de
  Laverny}}]{Recio-Blanco-06}
{Recio-Blanco}, A., {Bijaoui}, A., \& {de Laverny}, P. 2006, \mnras, 370, 141

\bibitem[{{Reid} {et~al.}(2014){Reid}, {Menten}, {Brunthaler}, {Zheng}, {Dame},
  {Xu}, {Wu}, {Zhang}, {Sanna}, {Sato}, {Hachisuka}, {Choi}, {Immer},
  {Moscadelli}, {Rygl}, \& {Bartkiewicz}}]{Reid-14}
{Reid}, M.~J., {Menten}, K.~M., {Brunthaler}, A., {et~al.} 2014, \apj, 783, 130

\bibitem[{{Ryabchikova} {et~al.}(2015){Ryabchikova}, {Piskunov}, {Kurucz},
  {Stempels}, {Heiter}, {Pakhomov}, \& {Barklem}}]{Ryabchikova-15}
{Ryabchikova}, T., {Piskunov}, N., {Kurucz}, R.~L., {et~al.} 2015, \physscr,
  90, 054005

\bibitem[{{Santos} {et~al.}(2004){Santos}, {Israelian}, \& {Mayor}}]{Santos-04}
{Santos}, N.~C., {Israelian}, G., \& {Mayor}, M. 2004, \aap, 415, 1153

\bibitem[{{Santos} {et~al.}(2011){Santos}, {Mayor}, {Bonfils}, {Dumusque},
  {Bouchy}, {Figueira}, {Lovis}, {Melo}, {Pepe}, {Queloz}, {S{\'e}gransan},
  {Sousa}, \& {Udry}}]{Santos-11}
{Santos}, N.~C., {Mayor}, M., {Bonfils}, X., {et~al.} 2011, \aap, 526, A112

\bibitem[{{Santos} {et~al.}(2007){Santos}, {Mayor}, {Bouchy}, {Pepe}, {Queloz},
  \& {Udry}}]{Santos-07}
{Santos}, N.~C., {Mayor}, M., {Bouchy}, F., {et~al.} 2007, \aap, 474, 647

\bibitem[{{Schilbach} {et~al.}(2006){Schilbach}, {Kharchenko}, {Piskunov},
  {R{\"o}ser}, \& {Scholz}}]{Schilbach-06}
{Schilbach}, E., {Kharchenko}, N.~V., {Piskunov}, A.~E., {R{\"o}ser}, S., \&
  {Scholz}, R.-D. 2006, \aap, 456, 523

\bibitem[{{Sch{\"o}nrich} {et~al.}(2010){Sch{\"o}nrich}, {Binney}, \&
  {Dehnen}}]{Schonrich-10}
{Sch{\"o}nrich}, R., {Binney}, J., \& {Dehnen}, W. 2010, \mnras, 403, 1829

\bibitem[{{Scott} {et~al.}(2006){Scott}, {Asplund}, {Grevesse}, \&
  {Sauval}}]{Scott-06}
{Scott}, P.~C., {Asplund}, M., {Grevesse}, N., \& {Sauval}, A.~J. 2006, \aap,
  456, 675

\bibitem[{{Sestito} \& {Randich}(2005)}]{Sestito-05}
{Sestito}, P. \& {Randich}, S. 2005, \aap, 442, 615

\bibitem[{{Slumstrup} {et~al.}(2017){Slumstrup}, {Grundahl}, {Brogaard},
  {Thygesen}, {Nissen}, {Jessen-Hansen}, {Van Eylen}, \&
  {Pedersen}}]{Slumstrup-17}
{Slumstrup}, D., {Grundahl}, F., {Brogaard}, K., {et~al.} 2017, \aap, 604, L8

\bibitem[{Smalley(2014)}]{Smalley-14}
Smalley, B. 2014, Stellar Parameters from Photometry, ed. E.~Niemczura,
  B.~Smalley, \& W.~Pych (Cham: Springer International Publishing), 111--120

\bibitem[{{Smiljanic} {et~al.}(2016){Smiljanic}, {Romano}, {Bragaglia},
  {Donati}, {Magrini}, {Friel}, {Jacobson}, {Randich}, {Ventura}, {Lind},
  {Bergemann}, {Nordlander}, {Morel}, {Pancino}, {Tautvai{\v s}ien{\.e}},
  {Adibekyan}, {Tosi}, {Vallenari}, {Gilmore}, {Bensby}, {Fran{\c c}ois},
  {Koposov}, {Lanzafame}, {Recio-Blanco}, {Bayo}, {Carraro}, {Casey},
  {Costado}, {Franciosini}, {Heiter}, {Hill}, {Hourihane}, {Jofr{\'e}},
  {Lardo}, {de Laverny}, {Lewis}, {Monaco}, {Morbidelli}, {Sacco}, {Sbordone},
  {Sousa}, {Worley}, \& {Zaggia}}]{Smiljanic-16}
{Smiljanic}, R., {Romano}, D., {Bragaglia}, A., {et~al.} 2016, \aap, 589, A115

\bibitem[{{Smith} {et~al.}(2001){Smith}, {Cunha}, \& {Lazzaro}}]{Smith-01}
{Smith}, V.~V., {Cunha}, K., \& {Lazzaro}, D. 2001, \aj, 121, 3207

\bibitem[{{Sneden} {et~al.}(1986){Sneden}, {Pilachowski}, \&
  {Vandenberg}}]{Sneden-86}
{Sneden}, C., {Pilachowski}, C.~A., \& {Vandenberg}, D.~A. 1986, \apj, 311, 826

\bibitem[{{Sneden}(1973)}]{Sneden-73}
{Sneden}, C.~A. 1973, PhD thesis, THE UNIVERSITY OF TEXAS AT AUSTIN.

\bibitem[{{Sousa}(2014)}]{Sousa-14}
{Sousa}, S.~G. 2014, [arXiv:1407.5817] [\eprint[arXiv]{1407.5817}]

\bibitem[{{Sousa} {et~al.}(2015{\natexlab{a}}){Sousa}, {Santos}, {Adibekyan},
  {Delgado-Mena}, \& {Israelian}}]{Sousa-15}
{Sousa}, S.~G., {Santos}, N.~C., {Adibekyan}, V., {Delgado-Mena}, E., \&
  {Israelian}, G. 2015{\natexlab{a}}, \aap, 577, A67

\bibitem[{{Sousa} {et~al.}(2008){Sousa}, {Santos}, {Mayor}, {Udry},
  {Casagrande}, {Israelian}, {Pepe}, {Queloz}, \& {Monteiro}}]{Sousa-08}
{Sousa}, S.~G., {Santos}, N.~C., {Mayor}, M., {et~al.} 2008, \aap, 487, 373

\bibitem[{{Sousa} {et~al.}(2015{\natexlab{b}}){Sousa}, {Santos}, {Mortier},
  {Tsantaki}, {Adibekyan}, {Delgado Mena}, {Israelian}, {Rojas-Ayala}, \&
  {Neves}}]{Sousa-15a}
{Sousa}, S.~G., {Santos}, N.~C., {Mortier}, A., {et~al.} 2015{\natexlab{b}},
  \aap, 576, A94

\bibitem[{{Sozzetti} {et~al.}(2007){Sozzetti}, {Torres}, {Charbonneau},
  {Latham}, {Holman}, {Winn}, {Laird}, \& {O'Donovan}}]{Sozzetti-07}
{Sozzetti}, A., {Torres}, G., {Charbonneau}, D., {et~al.} 2007, \apj, 664, 1190

\bibitem[{{Spina} {et~al.}(2016){Spina}, {Mel{\'e}ndez}, \&
  {Ram{\'{\i}}rez}}]{Spina-16}
{Spina}, L., {Mel{\'e}ndez}, J., \& {Ram{\'{\i}}rez}, I. 2016, \aap, 585, A152

\bibitem[{{Su{\'a}rez-Andr{\'e}s} {et~al.}(2017){Su{\'a}rez-Andr{\'e}s},
  {Israelian}, {Gonz{\'a}lez Hern{\'a}ndez}, {Adibekyan}, {Delgado Mena},
  {Santos}, \& {Sousa}}]{Suarez-17}
{Su{\'a}rez-Andr{\'e}s}, L., {Israelian}, G., {Gonz{\'a}lez Hern{\'a}ndez},
  J.~I., {et~al.} 2017, \aap, 599, A96

\bibitem[{{Su{\'a}rez Mascare{\~n}o} {et~al.}(2016){Su{\'a}rez Mascare{\~n}o},
  {Rebolo}, \& {Gonz{\'a}lez Hern{\'a}ndez}}]{Suarez-16}
{Su{\'a}rez Mascare{\~n}o}, A., {Rebolo}, R., \& {Gonz{\'a}lez Hern{\'a}ndez},
  J.~I. 2016, \aap, 595, A12

\bibitem[{{Su{\'a}rez Mascare{\~n}o} {et~al.}(2015){Su{\'a}rez Mascare{\~n}o},
  {Rebolo}, {Gonz{\'a}lez Hern{\'a}ndez}, \& {Esposito}}]{Suarez-15}
{Su{\'a}rez Mascare{\~n}o}, A., {Rebolo}, R., {Gonz{\'a}lez Hern{\'a}ndez},
  J.~I., \& {Esposito}, M. 2015, \mnras, 452, 2745

\bibitem[{{Tepfer} \& {Leach}(2006)}]{Tepfer-06}
{Tepfer}, D. \& {Leach}, S. 2006, \apss, 306, 69

\bibitem[{{Th{\'e}ado} \& {Vauclair}(2012)}]{Theado-12}
{Th{\'e}ado}, S. \& {Vauclair}, S. 2012, \apj, 744, 123

\bibitem[{{Torres}(2010)}]{Torres-10}
{Torres}, G. 2010, \aj, 140, 1158

\bibitem[{{Torres} {et~al.}(2012){Torres}, {Fischer}, {Sozzetti}, {Buchhave},
  {Winn}, {Holman}, \& {Carter}}]{Torres-12}
{Torres}, G., {Fischer}, D.~A., {Sozzetti}, A., {et~al.} 2012, \apj, 757, 161

\bibitem[{{Tsantaki} {et~al.}(2013){Tsantaki}, {Sousa}, {Adibekyan}, {Santos},
  {Mortier}, \& {Israelian}}]{Tsantaki-13}
{Tsantaki}, M., {Sousa}, S.~G., {Adibekyan}, V.~Z., {et~al.} 2013, \aap, 555,
  A150

\bibitem[{{Tsantaki} {et~al.}(2014){Tsantaki}, {Sousa}, {Santos}, {Montalto},
  {Delgado-Mena}, {Mortier}, {Adibekyan}, \& {Israelian}}]{Tsantaki-14}
{Tsantaki}, M., {Sousa}, S.~G., {Santos}, N.~C., {et~al.} 2014, \aap, 570, A80

\bibitem[{{Tucci Maia} {et~al.}(2016){Tucci Maia}, {Ram{\'{\i}}rez},
  {Mel{\'e}ndez}, {Bedell}, {Bean}, \& {Asplund}}]{TucciMaia-16}
{Tucci Maia}, M., {Ram{\'{\i}}rez}, I., {Mel{\'e}ndez}, J., {et~al.} 2016,
  \aap, 590, A32

\bibitem[{{van Leeuwen}(2007)}]{vanLeeuwen-07}
{van Leeuwen}, F. 2007, \aap, 474, 653

\bibitem[{{von Bloh} {et~al.}(2003){von Bloh}, {Franck}, {Bounama}, \&
  {Schellnhuber}}]{vonBloh-03}
{von Bloh}, W., {Franck}, S., {Bounama}, C., \& {Schellnhuber}, H.-J. 2003,
  Origins of Life and Evolution of the Biosphere, 33, 219

\bibitem[{{Weidner} \& {Kroupa}(2004)}]{Weidner-04}
{Weidner}, C. \& {Kroupa}, P. 2004, \mnras, 348, 187

\bibitem[{{Williams} \& {Gaidos}(2007)}]{Williams-07}
{Williams}, J.~P. \& {Gaidos}, E. 2007, \apjl, 663, L33

\bibitem[{{Wilson} \& {Matteucci}(1992)}]{Wilson-92}
{Wilson}, T.~L. \& {Matteucci}, F. 1992, \aapr, 4, 1

\bibitem[{{Wolff} {et~al.}(1985){Wolff}, {Heasley}, \& {Varsik}}]{Wolff-85}
{Wolff}, S.~C., {Heasley}, J.~N., \& {Varsik}, J. 1985, \pasp, 97, 707

\bibitem[{{Worley} {et~al.}(2016){Worley}, {de Laverny}, {Recio-Blanco},
  {Hill}, \& {Bijaoui}}]{Worley-16}
{Worley}, C.~C., {de Laverny}, P., {Recio-Blanco}, A., {Hill}, V., \&
  {Bijaoui}, A. 2016, \aap, 591, A81

\bibitem[{{Worley} {et~al.}(2012){Worley}, {de Laverny}, {Recio-Blanco},
  {Hill}, {Bijaoui}, \& {Ordenovic}}]{Worley-12}
{Worley}, C.~C., {de Laverny}, P., {Recio-Blanco}, A., {et~al.} 2012, \aap,
  542, A48

\bibitem[{{Wright} {et~al.}(2004){Wright}, {Marcy}, {Butler}, \&
  {Vogt}}]{Wright-04}
{Wright}, J.~T., {Marcy}, G.~W., {Butler}, R.~P., \& {Vogt}, S.~S. 2004, \apjs,
  152, 261

\bibitem[{{Yong} {et~al.}(2004){Yong}, {Lambert}, {Allende Prieto}, \&
  {Paulson}}]{Yong-04}
{Yong}, D., {Lambert}, D.~L., {Allende Prieto}, C., \& {Paulson}, D.~B. 2004,
  \apj, 603, 697

\end{thebibliography}

  \begin{appendix}
  
  \section{Tables}
  
\begin{table*}[t!]
\caption{\label{tab:params} Stellar parameters of the full sample stars and the S/N of the spectra.}
\centering
\begin{tabular}{llllllr}
\hline\hline
        Star &       \teff &     \logg$_{spec}$ &   \logg$_{gaia}$ &         [Fe/H] &          \vtur &   S/N \\
\hline
      HD631 &   6048$\pm$28 &  4.40$\pm$0.04 &  4.39$\pm$0.01 &   0.08$\pm$0.02 &  1.07$\pm$0.04 &    70 \\
     HD2247 &   6269$\pm$29 &  4.42$\pm$0.04 &  4.42$\pm$0.01 &   0.04$\pm$0.02 &  1.32$\pm$0.04 &   113 \\
     HD4021 &   5764$\pm$25 &  4.49$\pm$0.04 &  4.50$\pm$0.01 &   0.06$\pm$0.02 &  1.11$\pm$0.03 &   178 \\
     HD6204 &   5843$\pm$16 &  4.49$\pm$0.03 &  4.52$\pm$0.01 &   0.04$\pm$0.01 &  1.05$\pm$0.02 &   708 \\
     HD6245 &   5167$\pm$25 &  3.03$\pm$0.07 &  3.03$\pm$0.01 &   0.03$\pm$0.02 &  1.29$\pm$0.03 &   433 \\
     HD6790 &   6030$\pm$19 &  4.46$\pm$0.03 &  4.41$\pm$0.01 &   0.05$\pm$0.01 &  1.04$\pm$0.03 &   251 \\
     HD7515 &   5889$\pm$21 &  4.52$\pm$0.03 &  4.51$\pm$0.01 &  $-$0.04$\pm$0.02 &  1.05$\pm$0.03 &   206 \\
    HD10180 &   5913$\pm$14 &  4.38$\pm$0.02 &  4.35$\pm$0.01 &   0.10$\pm$0.01 &  1.09$\pm$0.02 &  1639 \\
    HD10678 &   5609$\pm$20 &  4.38$\pm$0.04 &  4.50$\pm$0.01 &   0.10$\pm$0.02 &  0.88$\pm$0.03 &    69 \\
    HD22249 &   5737$\pm$20 &  4.54$\pm$0.04 &  4.55$\pm$0.01 &  $-$0.04$\pm$0.02 &  0.90$\pm$0.03 &   175 \\
    HD30858 &   5109$\pm$45 &  4.33$\pm$0.09 &  4.50$\pm$0.02 &  $-$0.07$\pm$0.03 &  0.66$\pm$0.10 &    84 \\
    HD37811 &   5146$\pm$31 &  2.91$\pm$0.08 &  2.78$\pm$0.01 &   0.01$\pm$0.03 &  1.55$\pm$0.03 &   307 \\
    HD38382 &   6076$\pm$13 &  4.45$\pm$0.03 &  4.37$\pm$0.01 &   0.04$\pm$0.01 &  1.19$\pm$0.02 &  1302 \\
    HD41071 &   5593$\pm$40 &  4.64$\pm$0.05 &  4.58$\pm$0.01 &   0.01$\pm$0.03 &  1.29$\pm$0.06 &   116 \\
    HD41842 &   5012$\pm$51 &  4.36$\pm$0.12 &  4.57$\pm$0.02 &  $-$0.02$\pm$0.03 &  0.81$\pm$0.12 &   201 \\
    HD45415 &   4830$\pm$47 &  2.67$\pm$0.10 &  2.60$\pm$0.04 &  $-$0.04$\pm$0.04 &  1.47$\pm$0.05 &   215 \\
    HD52456 &   5110$\pm$46 &  4.33$\pm$0.09 &  4.51$\pm$0.02 &  $-$0.02$\pm$0.03 &  0.65$\pm$0.11 &   108 \\
    HD56351 &   5734$\pm$19 &  4.63$\pm$0.02 &  4.56$\pm$0.01 &  $-$0.07$\pm$0.02 &  1.17$\pm$0.00 &   114 \\
    HD62412 &   4958$\pm$42 &  2.71$\pm$0.08 &  2.77$\pm$0.02 &   0.04$\pm$0.04 &  1.45$\pm$0.04 &   323 \\
    HD62816 &   6553$\pm$57 &  4.75$\pm$0.08 &  4.39$\pm$0.02 &   0.09$\pm$0.04 &  1.67$\pm$0.08 &   276 \\
    HD64942 &   5875$\pm$22 &  4.56$\pm$0.04 &  4.52$\pm$0.01 &   0.04$\pm$0.02 &  1.18$\pm$0.03 &   196 \\
    HD74006 &   5304$\pm$64 &  2.76$\pm$0.10 &  2.50$\pm$0.03 &   0.07$\pm$0.06 &  2.57$\pm$0.11 &   755 \\
    HD77191 &   5785$\pm$40 &  4.50$\pm$0.07 &  4.51$\pm$0.02 &  $-$0.02$\pm$0.03 &  1.19$\pm$0.06 &   180 \\
    HD89124 &   5668$\pm$18 &  4.45$\pm$0.03 &  4.51$\pm$0.01 &  $-$0.04$\pm$0.02 &  0.85$\pm$0.03 &    69 \\
    HD89839 &   6290$\pm$20 &  4.44$\pm$0.03 &  4.32$\pm$0.01 &   0.05$\pm$0.02 &  1.35$\pm$0.03 &   576 \\
    HD89965 &   4939$\pm$68 &  4.32$\pm$0.15 &  4.56$\pm$0.03 &  $-$0.08$\pm$0.04 &  0.53$\pm$0.21 &   318 \\
    HD92987 &   5860$\pm$15 &  4.18$\pm$0.02 &  4.12$\pm$0.01 &   0.07$\pm$0.01 &  1.16$\pm$0.02 &   424 \\
    HD95542 &   5971$\pm$17 &  4.48$\pm$0.03 &  4.48$\pm$0.01 &  $-$0.02$\pm$0.01 &  1.06$\pm$0.03 &   518 \\
    HD96116 &   5841$\pm$14 &  4.49$\pm$0.02 &  4.52$\pm$0.01 &   0.02$\pm$0.01 &  0.97$\pm$0.02 &   609 \\
    HD99648 &   5066$\pm$50 &  2.60$\pm$0.15 &  2.36$\pm$0.02 &   0.06$\pm$0.05 &  1.82$\pm$0.06 &   832 \\
   HD109098 &   5897$\pm$16 &  4.10$\pm$0.02 &  4.04$\pm$0.01 &   0.10$\pm$0.01 &  1.23$\pm$0.02 &   193 \\
   HD115341 &   6008$\pm$17 &  4.44$\pm$0.02 &  4.41$\pm$0.01 &   0.00$\pm$0.01 &  1.07$\pm$0.02 &   166 \\
   HD118563 &   5481$\pm$19 &  4.44$\pm$0.04 &  4.51$\pm$0.01 &  $-$0.03$\pm$0.01 &  0.79$\pm$0.03 &   204 \\
   HD126829 &  4535$\pm$108 &  4.28$\pm$0.30 &  4.58$\pm$0.04 &  $-$0.04$\pm$0.04 &  0.58$\pm$0.40 &   184 \\
   HD131218 &   5743$\pm$17 &  4.44$\pm$0.03 &  4.50$\pm$0.01 &  $-$0.02$\pm$0.01 &  0.91$\pm$0.03 &   107 \\
   HD136894 &   5406$\pm$16 &  4.34$\pm$0.03 &  4.45$\pm$0.01 &  $-$0.09$\pm$0.01 &  0.75$\pm$0.03 &  1204 \\
   HD155717 &   4942$\pm$55 &  4.46$\pm$0.13 &  4.60$\pm$0.02 &  $-$0.08$\pm$0.03 &  0.77$\pm$0.12 &   119 \\
   HD158469 &   6190$\pm$19 &  4.36$\pm$0.03 &  4.27$\pm$0.01 &   0.06$\pm$0.01 &  1.32$\pm$0.02 &   429 \\
   HD167554 &   5291$\pm$28 &  4.49$\pm$0.05 &  4.62$\pm$0.01 &  $-$0.07$\pm$0.02 &  0.90$\pm$0.06 &   270 \\
   HD176535 &   4635$\pm$46 &  4.38$\pm$0.18 &  4.57$\pm$0.02 &  $-$0.05$\pm$0.14 &  0.02$\pm$3.52 &   245 \\
   HD181387 &   6367$\pm$44 &  4.22$\pm$0.05 &  4.28$\pm$0.01 &   0.02$\pm$0.03 &  1.16$\pm$0.09 &   189 \\
   HD181517 &   4999$\pm$41 &  2.87$\pm$0.11 &  2.87$\pm$0.03 &   0.07$\pm$0.03 &  1.43$\pm$0.04 &   330 \\
   HD186302 &   5675$\pm$15 &  4.43$\pm$0.02 &  4.47$\pm$0.01 &   0.00$\pm$0.01 &  0.86$\pm$0.03 &   159 \\
   HD190204 &   5406$\pm$23 &  4.46$\pm$0.04 &  4.57$\pm$0.01 &   0.00$\pm$0.02 &  1.03$\pm$0.04 &   181 \\
   HD197300 &   5939$\pm$24 &  4.52$\pm$0.03 &  4.52$\pm$0.01 &   0.01$\pm$0.02 &  1.16$\pm$0.03 &   107 \\
   HD199951 &   5272$\pm$37 &  3.11$\pm$0.07 &  2.91$\pm$0.02 &   0.03$\pm$0.03 &  1.59$\pm$0.03 &   722 \\
   HD203387 &   5288$\pm$28 &  3.12$\pm$0.08 &  2.87$\pm$0.01 &   0.09$\pm$0.03 &  1.52$\pm$0.03 &   880 \\
   HD209458 &   6139$\pm$21 &  4.47$\pm$0.03 &  4.39$\pm$0.01 &   0.06$\pm$0.02 &  1.21$\pm$0.03 &  1114 \\
   HD212563 &   4946$\pm$53 &  4.37$\pm$0.15 &  4.57$\pm$0.02 &   0.00$\pm$0.03 &  0.71$\pm$0.13 &   124 \\
   HD216530 &   4750$\pm$84 &  4.27$\pm$0.22 &  4.55$\pm$0.03 &   0.02$\pm$0.04 &  0.74$\pm$0.22 &    85 \\
   HD218614 &   5666$\pm$20 &  4.48$\pm$0.03 &  4.52$\pm$0.01 &  $-$0.03$\pm$0.02 &  1.02$\pm$0.03 &   114 \\
   HIP96240 &   4767$\pm$55 &  4.25$\pm$0.15 &  4.56$\pm$0.02 &  $-$0.10$\pm$0.02 &  0.63$\pm$0.14 &   119 \\
 BD+004175B &   6354$\pm$50 &  4.45$\pm$0.05 &  4.41$\pm$0.02 &   0.08$\pm$0.04 &  1.46$\pm$0.07 &    77 \\
    CoRoT-7 &   5309$\pm$37 &  4.42$\pm$0.09 &  4.56$\pm$0.02 &   0.02$\pm$0.03 &  0.96$\pm$0.07 &   566 \\
\hline
\end{tabular}
\end{table*}

\begin{table*}[t!]
\caption{\label{tab:age} Stellar mass, radius, age, and activity index for the full sample.}
\centering
\begin{tabular}{lllll}
\hline\hline  
        Star &          M$_{star}$ &           R$_{star}$ &  Age$_{padova}$ &       $\log(R'_{HK}$) \\
\hline
       HD631 &  1.14$\pm$0.02 &   1.11$\pm$0.02 &   1.87$\pm$0.87 &  $-$4.93$\pm$0.06 \\
      HD2247 &  1.21$\pm$0.02 &   1.16$\pm$0.03 &   0.65$\pm$0.50 &  $-$5.29$\pm$0.13 \\
      HD4021 &  1.04$\pm$0.02 &   0.95$\pm$0.01 &   1.02$\pm$0.80 &  $-$4.36$\pm$0.03 \\
      HD6204 &  1.05$\pm$0.01 &   0.95$\pm$0.01 &   0.29$\pm$0.17 &  $-$4.49$\pm$0.03 \\
      HD6245 &  2.34$\pm$0.02 &   7.53$\pm$0.19 &   0.77$\pm$0.02 &  $-$4.84$\pm$0.05 \\
      HD6790 &  1.12$\pm$0.01 &   1.09$\pm$0.02 &   1.81$\pm$0.61 &  $-$4.80$\pm$0.01 \\
      HD7515 &  1.04$\pm$0.01 &   0.95$\pm$0.02 &   0.54$\pm$0.42 &  $-$4.42$\pm$0.02 \\
     HD10180 &  1.08$\pm$0.01 &   1.15$\pm$0.00 &   4.44$\pm$0.34 &  $-$4.92$\pm$0.01 \\
     HD10678 &  1.00$\pm$0.02 &   0.91$\pm$0.01 &   2.19$\pm$1.24 &  $-$4.67$\pm$0.01 \\
     HD22249 &  1.00$\pm$0.01 &   0.87$\pm$0.01 &   0.54$\pm$0.40 &  $-$4.49$\pm$0.02 \\
     HD30858 &  0.82$\pm$0.02 &   0.79$\pm$0.01 &   8.40$\pm$3.07 &  $-$4.60$\pm$0.08 \\
     HD37811 &  2.83$\pm$0.03 &  10.93$\pm$0.24 &   0.46$\pm$0.01 &  $-$5.28$\pm$0.01 \\
     HD38382 &  1.13$\pm$0.01 &   1.15$\pm$0.00 &   2.68$\pm$0.29 &  $-$4.89$\pm$0.02 \\
     HD41071 &  0.95$\pm$0.02 &   0.84$\pm$0.02 &   0.87$\pm$0.81 &  $-$4.31$\pm$0.02 \\
     HD41842 &  0.81$\pm$0.02 &   0.75$\pm$0.01 &   4.67$\pm$3.67 &  $-$4.37$\pm$0.09 \\
     HD45415 &  1.65$\pm$0.15 &  10.45$\pm$0.36 &   2.22$\pm$0.55 &  $-$5.43$\pm$0.01 \\
     HD52456 &  0.83$\pm$0.02 &   0.80$\pm$0.01 &   7.45$\pm$3.62 &  $-$4.72$\pm$0.01 \\
     HD56351 &  0.98$\pm$0.01 &   0.87$\pm$0.00 &   0.57$\pm$0.47 &  $-$4.35$\pm$0.04 \\
     HD62412 &  2.34$\pm$0.09 &   10.4$\pm$0.20 &   0.87$\pm$0.13 &  $-$5.20$\pm$0.03 \\
     HD62816 &  1.30$\pm$0.02 &   1.25$\pm$0.02 &   0.35$\pm$0.22 &  $-$4.77$\pm$0.03 \\
     HD64942 &  1.06$\pm$0.01 &   0.96$\pm$0.00 &   0.36$\pm$0.24 &  $-$4.42$\pm$0.04 \\
     HD74006 &  3.45$\pm$0.12 &  16.58$\pm$1.23 &   0.27$\pm$0.02 &  $-$5.06$\pm$0.02 \\
     HD77191 &  1.01$\pm$0.02 &   0.93$\pm$0.02 &   1.46$\pm$1.26 &  $-$4.37$\pm$0.01 \\
     HD89124 &  0.97$\pm$0.02 &   0.91$\pm$0.01 &   2.60$\pm$1.28 &  $-$4.62$\pm$0.04 \\
     HD89839 &  1.22$\pm$0.01 &   1.26$\pm$0.00 &   1.95$\pm$0.27 &  $-$4.98$\pm$0.10 \\
     HD89965 &  0.79$\pm$0.02 &   0.73$\pm$0.01 &   5.09$\pm$4.06 &  $-$4.66$\pm$0.04 \\
     HD92987 &  1.10$\pm$0.01 &   1.51$\pm$0.01 &   6.95$\pm$0.16 &  $-$4.89$\pm$0.00 \\
     HD95542 &  1.08$\pm$0.01 &   1.00$\pm$0.01 &   0.67$\pm$0.48 &  $-$4.64$\pm$0.06 \\
     HD96116 &  1.05$\pm$0.01 &   0.95$\pm$0.01 &   0.33$\pm$0.21 &  $-$4.61$\pm$0.04 \\
     HD99648 &  3.66$\pm$0.11 &  19.92$\pm$1.38 &   0.23$\pm$0.02 &  $-$4.77$\pm$0.00 \\
    HD109098 &  1.18$\pm$0.01 &   1.69$\pm$0.04 &   5.52$\pm$0.23 &  $-$4.98$\pm$0.05 \\
    HD115341 &  1.09$\pm$0.01 &   1.07$\pm$0.03 &   2.67$\pm$0.52 &  $-$4.90$\pm$0.15 \\
    HD118563 &  0.91$\pm$0.02 &   0.87$\pm$0.01 &   4.79$\pm$1.76 &  $-$4.69$\pm$0.02 \\
    HD126829 &  0.71$\pm$0.01 &   0.66$\pm$0.01 &   4.83$\pm$3.98 &  $-$4.44$\pm$0.02 \\
    HD131218 &  1.00$\pm$0.02 &   0.91$\pm$0.01 &   1.85$\pm$1.06 &  $-$4.68$\pm$0.15 \\
    HD136894 &  0.85$\pm$0.01 &   0.87$\pm$0.01 &  11.08$\pm$0.55 &  $-$4.93$\pm$0.01 \\
    HD155717 &  0.77$\pm$0.02 &   0.72$\pm$0.01 &   4.50$\pm$3.93 &  $-$4.52$\pm$0.03 \\
    HD158469 &  1.20$\pm$0.00 &   1.32$\pm$0.01 &   3.03$\pm$0.18 &  $-$4.95$\pm$0.02 \\
    HD167554 &  0.86$\pm$0.01 &   0.76$\pm$0.00 &   0.71$\pm$0.68 &  $-$4.34$\pm$0.02 \\
    HD176535 &  0.75$\pm$0.02 &   0.69$\pm$0.02 &   5.64$\pm$4.10 &  $-$4.81$\pm$0.04 \\
    HD181387 &  1.25$\pm$0.02 &   1.34$\pm$0.03 &   2.00$\pm$0.41 &  $-$5.06$\pm$0.04 \\
    HD181517 &  2.33$\pm$0.13 &   9.33$\pm$0.31 &   0.84$\pm$0.16 &  $-$5.38$\pm$0.06 \\
    HD186302 &  0.97$\pm$0.01 &   0.95$\pm$0.01 &   4.50$\pm$0.81 &  $-$4.84$\pm$0.06 \\
    HD190204 &  0.92$\pm$0.02 &   0.83$\pm$0.02 &   1.31$\pm$1.16 &  $-$4.37$\pm$0.03 \\
    HD197300 &  1.07$\pm$0.01 &   0.96$\pm$0.00 &   0.37$\pm$0.25 &  $-$4.36$\pm$0.02 \\
    HD199951 &  2.57$\pm$0.05 &   9.03$\pm$0.32 &   0.60$\pm$0.03 &  $-$4.65$\pm$0.03 \\
    HD203387 &  2.67$\pm$0.03 &   9.64$\pm$0.29 &   0.55$\pm$0.02 &  $-$4.72$\pm$0.01 \\
    HD209458 &  1.17$\pm$0.02 &   1.14$\pm$0.02 &   1.42$\pm$0.56 &  $-$4.91$\pm$0.01 \\
    HD212563 &  0.80$\pm$0.02 &   0.75$\pm$0.02 &   4.83$\pm$3.83 &  $-$4.31$\pm$0.01 \\
    HD216530 &  0.77$\pm$0.02 &   0.72$\pm$0.01 &   5.00$\pm$3.95 &  $-$4.33$\pm$0.00 \\
    HD218614 &  0.98$\pm$0.02 &   0.89$\pm$0.02 &   1.81$\pm$1.19 &  $-$4.37$\pm$0.00 \\
    HIP96240 &  0.75$\pm$0.02 &   0.70$\pm$0.01 &   5.97$\pm$4.19 &  $-$4.54$\pm$0.03 \\
  BD+004175B &  1.22$\pm$0.02 &   1.16$\pm$0.02 &   0.56$\pm$0.42 &  $-$5.02$\pm$0.09 \\
     CoRoT-7 &  0.89$\pm$0.02 &   0.81$\pm$0.02 &   2.93$\pm$2.64 &  $-$4.82$\pm$0.07 \\
 NAME TrES-1 &  0.88$\pm$0.01 &   0.91$\pm$0.01 &  10.15$\pm$1.37 &  $-$4.79$\pm$0.00 \\
\hline
\end{tabular}
\end{table*}

\begin{table*}[t!]
\caption{\label{tab:kinematics} Kinematic parameters of the full sample stars.}
\centering
\begin{tabular}{lrrrrrlllll}
\hline\hline  
        Star &  RV  &     U$_{LSR}$ &     V$_{LSR}$ &     W$_{LSR}$ &      Z &      R &    $e$ &   Z$_{max}$ &    R$_{per}$ &    R$_{ap}$ \\
\hline
       HD631  & 31.9 &   $-$5.5  & 23.1 &  $-$22.0  &  $-$0.026  & 8.354 & 0.122 & 0.401 & 10.662 &   8.342 \\
      HD2247  & 20.6 & 6.4 & 4.8 &  $-$14.7  &  $-$0.115  & 8.336 & 0.034 & 0.261 & 8.860 &   8.275 \\
      HD4021  &           $-$3.0  & 11.2 & 15.5 & 10.8 &  $-$0.034  & 8.344 & 0.088 & 0.178 & 9.872 &   8.275 \\
      HD6204  &           $-$3.0  & 8.7 & 14.6 & 11.4 &  $-$0.026  & 8.351 & 0.081 & 0.184 & 9.769 &   8.306 \\
      HD6245  &           $-$1.5  & 9.9 & 16.5 & 7.2 &  $-$0.046  & 8.329 & 0.090 & 0.125 & 9.923 &   8.277 \\
      HD6790  &          $-$23.8  &   $-$1.5  & 20.2 & 29.3 &  $-$0.071  & 8.335 & 0.109 & 0.566 & 10.375 &   8.334 \\
      HD7515  & 1.0 & 24.2 & 5.4 & 10.9 &  $-$0.050  & 8.327 & 0.087 & 0.177 & 9.374 &   7.871 \\
     HD10180  & 35.4 & 20.4 &   $-$3.7  &  $-$23.3  &  $-$0.007  & 8.332 & 0.071 & 0.379 & 8.832 &   7.666 \\
     HD10678  & 16.0 & 15.9 &   $-$3.5  &   $-$1.4  &  $-$0.013  & 8.332 & 0.056 & 0.025 & 8.673 &   7.746 \\
     HD22249  & 15.9 & 4.9 &   $-$3.7  & 0.3 &  $-$0.017  & 8.348 & 0.025 & 0.017 & 8.402 &   7.990 \\
     HD30858  & 25.9 &   $-$0.6  & 1.2 &  $-$13.8  &  $-$0.006  & 8.368 & 0.008 & 0.204 & 8.506 &   8.366 \\
     HD37811  &           $-$5.6  & 31.5 & 13.8 &   $-$1.9  &  $-$0.030  & 8.395 & 0.129 & 0.048 & 10.301 &   7.951 \\
     HD38382  & 37.8 &  $-$16.8  &   $-$8.3  &   $-$8.0  & 0.015 & 8.357 & 0.070 & 0.117 & 8.612 &   7.488 \\
     HD41071  & 24.9 &   $-$0.1  &   $-$9.8  & 1.0 & 0.000 & 8.355 & 0.049 & 0.014 & 8.355 &   7.572 \\
     HD41842  & 12.4 & 0.0 & 6.8 & 2.2 & 0.012 & 8.360 & 0.035 & 0.034 & 8.961 &   8.360 \\
     HD45415  & 52.7 &  $-$39.7  & 0.4 &  $-$10.4  & 0.019 & 8.421 & 0.135 & 0.169 & 9.749 &   7.434 \\
     HD52456  &          $-$12.0  & 19.5 & 22.3 & 4.1 & 0.028 & 8.365 & 0.131 & 0.077 & 10.710 &   8.225 \\
     HD56351  &           $-$5.0  & 18.8 & 16.2 & 9.3 & 0.008 & 8.344 & 0.105 & 0.151 & 10.090 &   8.173 \\
     HD62412  &          $-$16.1  & 31.7 & 20.8 &   $-$0.9  & 0.023 & 8.388 & 0.152 & 0.031 & 10.918 &   8.043 \\
     HD62816  &           $-$4.5  & 16.5 & 14.7 & 8.5 & 0.032 & 8.378 & 0.094 & 0.142 & 9.942 &   8.230 \\
     HD64942  &           $-$8.4  & 23.4 & 13.7 & 7.2 & 0.033 & 8.372 & 0.106 & 0.122 & 10.017 &   8.091 \\
     HD74006  &          $-$11.4  & 28.7 & 19.9 & 4.0 & 0.032 & 8.366 & 0.141 & 0.078 & 10.707 &   8.064 \\
     HD77191  &           $-$7.5  & 14.2 & 17.0 & 0.8 & 0.059 & 8.380 & 0.099 & 0.069 & 10.106 &   8.280 \\
     HD89124  &           $-$4.0  & 17.4 & 18.3 & 7.7 & 0.012 & 8.319 & 0.110 & 0.126 & 10.206 &   8.184 \\
     HD89839  & 31.9 & 15.3 &  $-$19.7  & 11.3 & 0.028 & 8.329 & 0.111 & 0.167 & 8.442 &   6.759 \\
     HD89965  &           $-$7.0  & 8.1 & 20.6 & 8.5 & 0.042 & 8.342 & 0.108 & 0.149 & 10.331 &   8.314 \\
     HD92987  & 4.5 & 13.0 & 9.7 & 14.1 & 0.038 & 8.335 & 0.067 & 0.227 & 9.384 &   8.203 \\
     HD95542  & 15.4 & 15.3 &   $-$3.7  & 9.6 & 0.019 & 8.320 & 0.055 & 0.142 & 8.640 &   7.745 \\
     HD96116  & 31.3 & 31.1 &  $-$14.9  & 0.5 & 0.027 & 8.321 & 0.128 & 0.029 & 8.816 &   6.812 \\
     HD99648  &           $-$8.8  & 28.8 & 14.5 & 0.1 & 0.159 & 8.354 & 0.124 & 0.186 & 10.231 &   7.981 \\
    HD109098  & 2.0 & 1.3 & 15.4 & 13.5 & 0.075 & 8.329 & 0.079 & 0.236 & 9.766 &   8.328 \\
    HD115341  & 19.4 & 21.5 &   $-$4.7  & 18.0 & 0.065 & 8.322 & 0.076 & 0.292 & 8.806 &   7.569 \\
    HD118563  &          $-$18.9  &   $-$8.7  & 17.3 & 1.1 & 0.047 & 8.305 & 0.092 & 0.056 & 9.941 &   8.266 \\
    HD126829  &          $-$18.4  &   $-$3.6  & 15.8 &   $-$3.4  & 0.048 & 8.316 & 0.081 & 0.075 & 9.769 &   8.309 \\
    HD131218  & 24.9 & 23.4 &  $-$11.0  & 3.8 & 0.008 & 8.306 & 0.096 & 0.055 & 8.671 &   7.151 \\
    HD136894  &           $-$6.4  & 6.3 & 10.2 & 0.6 & 0.036 & 8.317 & 0.055 & 0.040 & 9.255 &   8.283 \\
    HD155717  & 12.5 & 20.5 & 14.8 & 15.8 & 0.043 & 8.297 & 0.103 & 0.267 & 9.951 &   8.087 \\
    HD158469  & 39.3 & 50.8 & 5.2 & 18.3 & 0.021 & 8.274 & 0.174 & 0.319 & 10.268 &   7.220 \\
    HD167554  & 21.9 & 32.4 & 13.7 &   $-$0.2  & 0.014 & 8.283 & 0.130 & 0.017 & 10.154 &   7.821 \\
    HD176535  &          $-$31.9  &  $-$15.9  &   $-$5.5  & 13.6 & 0.020 & 8.306 & 0.060 & 0.203 & 8.596 &   7.623 \\
    HD181387  &          $-$32.0  &   $-$2.8  &  $-$25.4  & 8.2 & 0.009 & 8.221 & 0.129 & 0.116 & 8.225 &   6.348 \\
    HD181517  & 10.0 & 15.6 & 8.9 &   $-$6.7  &  $-$0.040  & 8.215 & 0.069 & 0.110 & 9.204 &   8.020 \\
    HD186302  &           $-$1.0  & 8.4 & 2.5 & 13.8 &  $-$0.004  & 8.300 & 0.032 & 0.206 & 8.688 &   8.154 \\
    HD190204  & 13.0 & 25.0 & 22.0 & 4.4 &  $-$0.009  & 8.285 & 0.139 & 0.073 & 10.677 &   8.064 \\
    HD197300  &          $-$13.9  &   $-$7.9  & 7.3 & 4.2 &  $-$0.009  & 8.294 & 0.045 & 0.062 & 9.002 &   8.223 \\
    HD199951  & 19.0 & 27.8 & 10.9 &   $-$2.9  &  $-$0.019  & 8.289 & 0.109 & 0.051 & 9.801 &   7.872 \\
    HD203387  & 12.4 & 11.2 & 18.9 &   $-$6.6  &  $-$0.015  & 8.301 & 0.103 & 0.107 & 10.133 &   8.244 \\
    HD209458  &          $-$15.0  & 4.0 &   $-$3.6  & 7.7 & 0.002 & 8.330 & 0.022 & 0.110 & 8.370 &   8.002 \\
    HD212563  &          $-$13.4  & 2.3 & 1.0 & 14.9 &  $-$0.012  & 8.319 & 0.010 & 0.221 & 8.461 &   8.289 \\
    HD216530  & 14.0 & 16.0 & 16.9 &   $-$6.4  &  $-$0.016  & 8.318 & 0.101 & 0.104 & 10.044 &   8.193 \\
    HD218614  & 7.9 & 19.8 & 22.0 & 5.2 &  $-$0.022  & 8.327 & 0.130 & 0.090 & 10.633 &   8.182 \\
    HIP96240  & 5.0 & 12.4 & 22.2 & 4.7 & 0.009 & 8.295 & 0.119 & 0.077 & 10.465 &   8.235 \\
  BD+004175B  &          $-$47.0  &  $-$14.6  &  $-$34.5  & 9.8 & 0.009 & 8.221 & 0.180 & 0.140 & 8.283 &   5.759 \\
     CoRoT-7  & 31.2 &  $-$11.0  &   $-$8.0  & 13.5 & 0.019 & 8.474 & 0.053 & 0.204 & 8.611 &   7.748 \\
 NAME TrES-1  &          $-$20.7  & 20.7 &  $-$18.6  & 18.0 & 0.062 & 8.282 & 0.115 & 0.284 & 8.491 &   6.744 \\
\hline
\end{tabular}
\end{table*}

\section{Comparison of stellar parameters}

\begin{figure*}
\begin{center}
\begin{tabular}{cc}
\includegraphics[angle=0,width=0.5\linewidth]{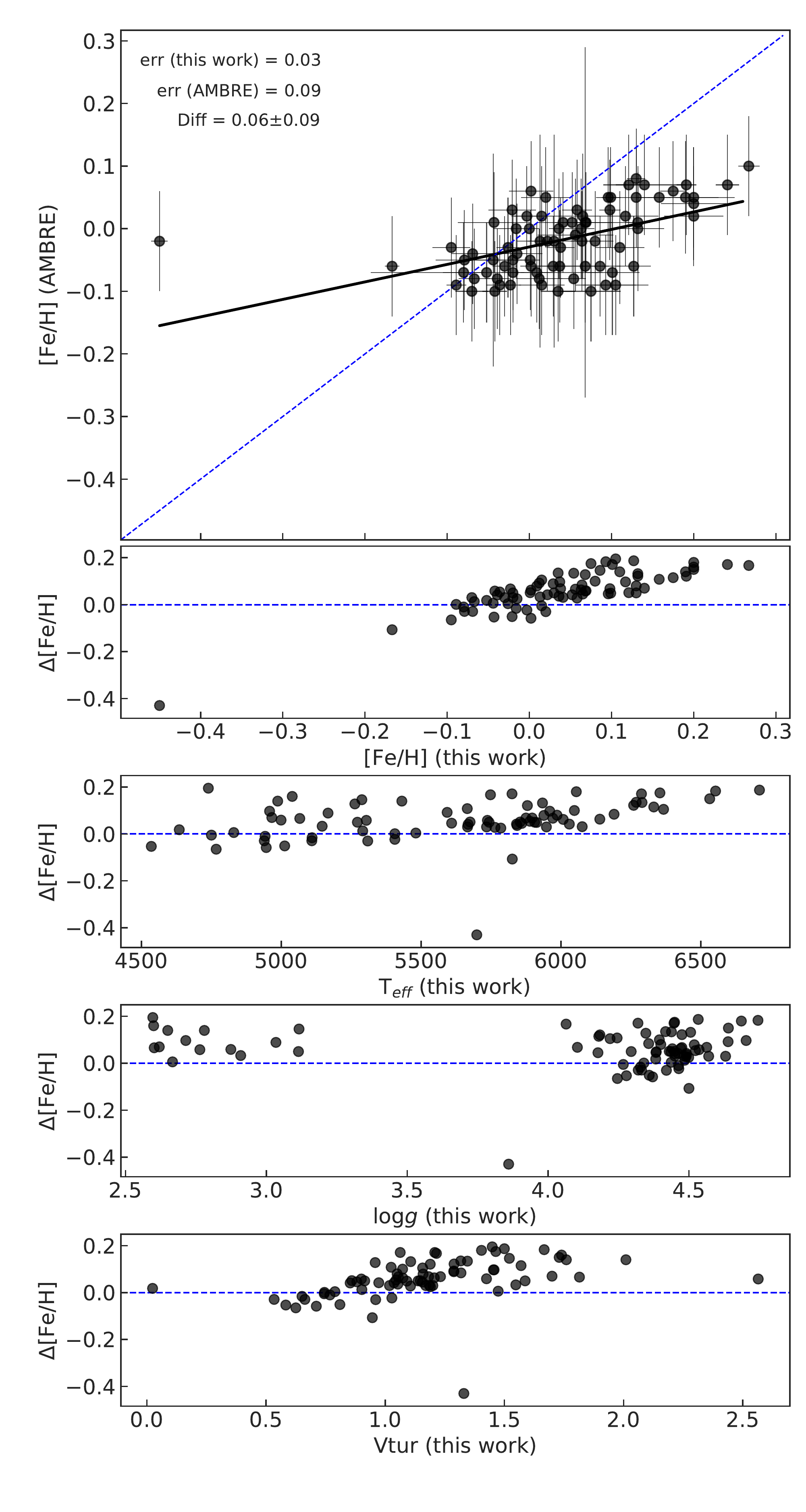}
\includegraphics[angle=0,width=0.5\linewidth]{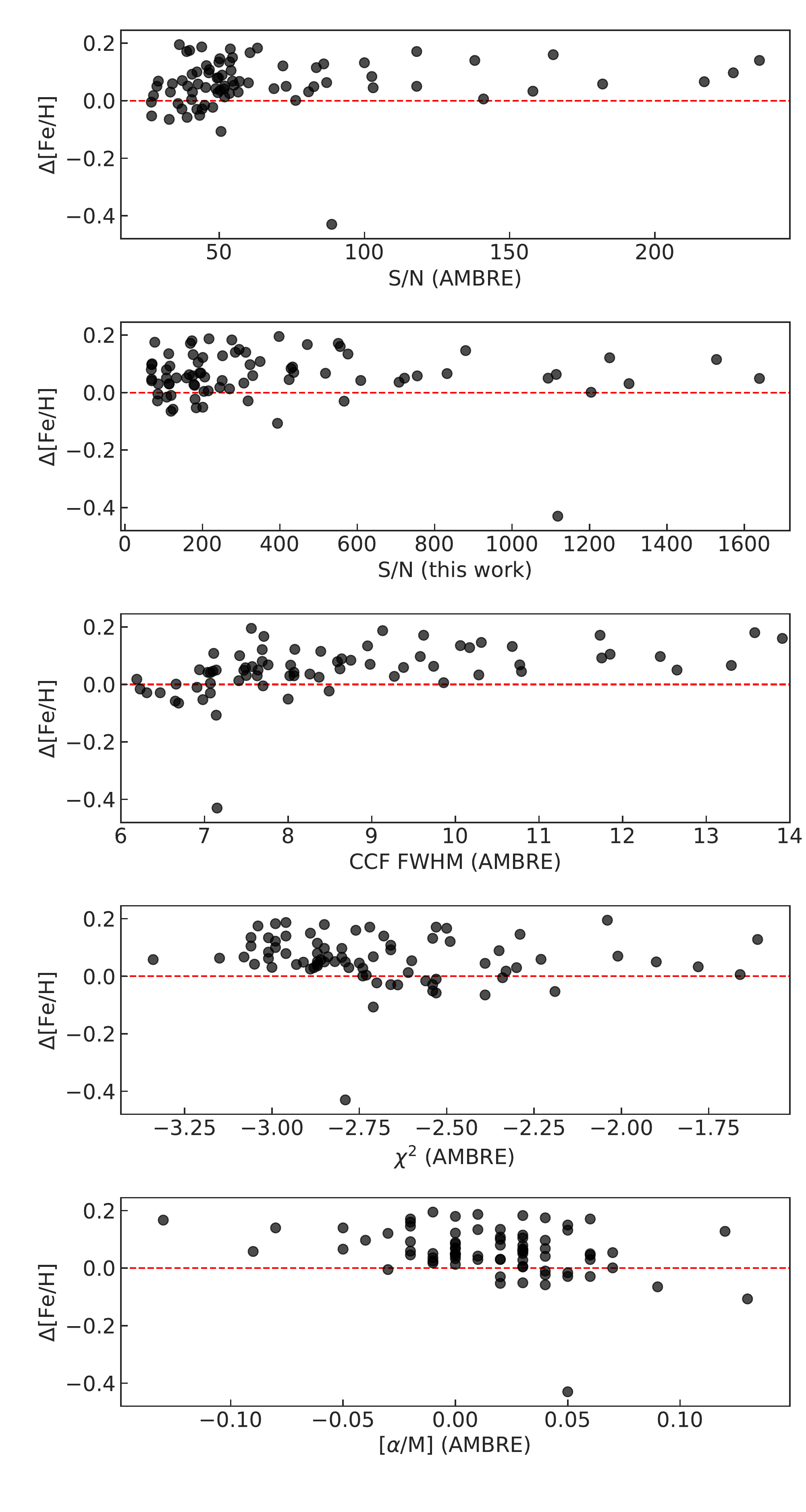}
\end{tabular}
\end{center}
\vspace{-0.6cm}
\caption{Comparison of the stellar mean metallicities derived within the AMBRE project and [Fe/H] derived here.}
\label{fig_feh_ambre_thiswork}
\end{figure*}

\begin{figure*}
\begin{center}
\begin{tabular}{ccc}
\includegraphics[angle=0,width=0.37\linewidth]{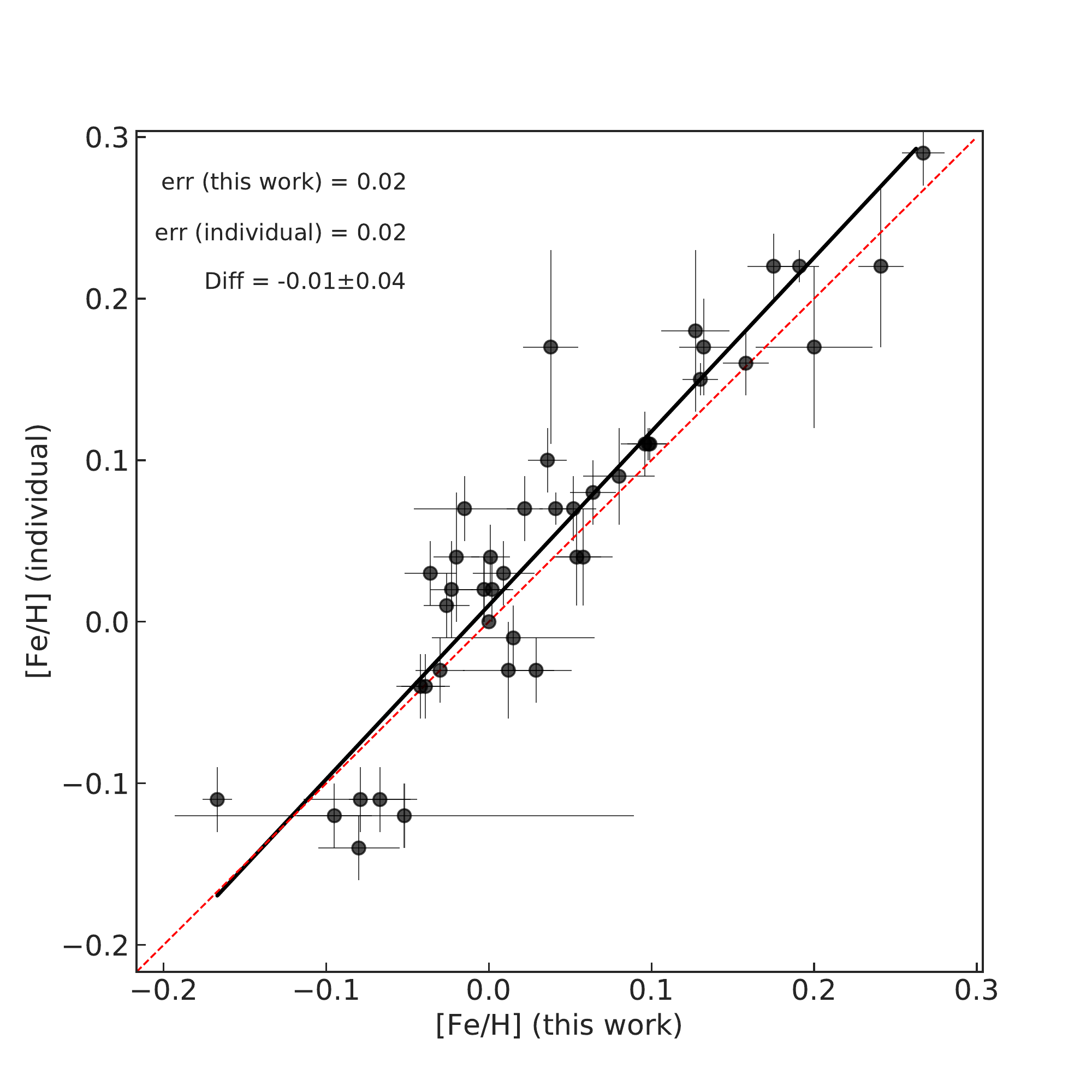} \hspace{-0.9cm}
\includegraphics[angle=0,width=0.37\linewidth]{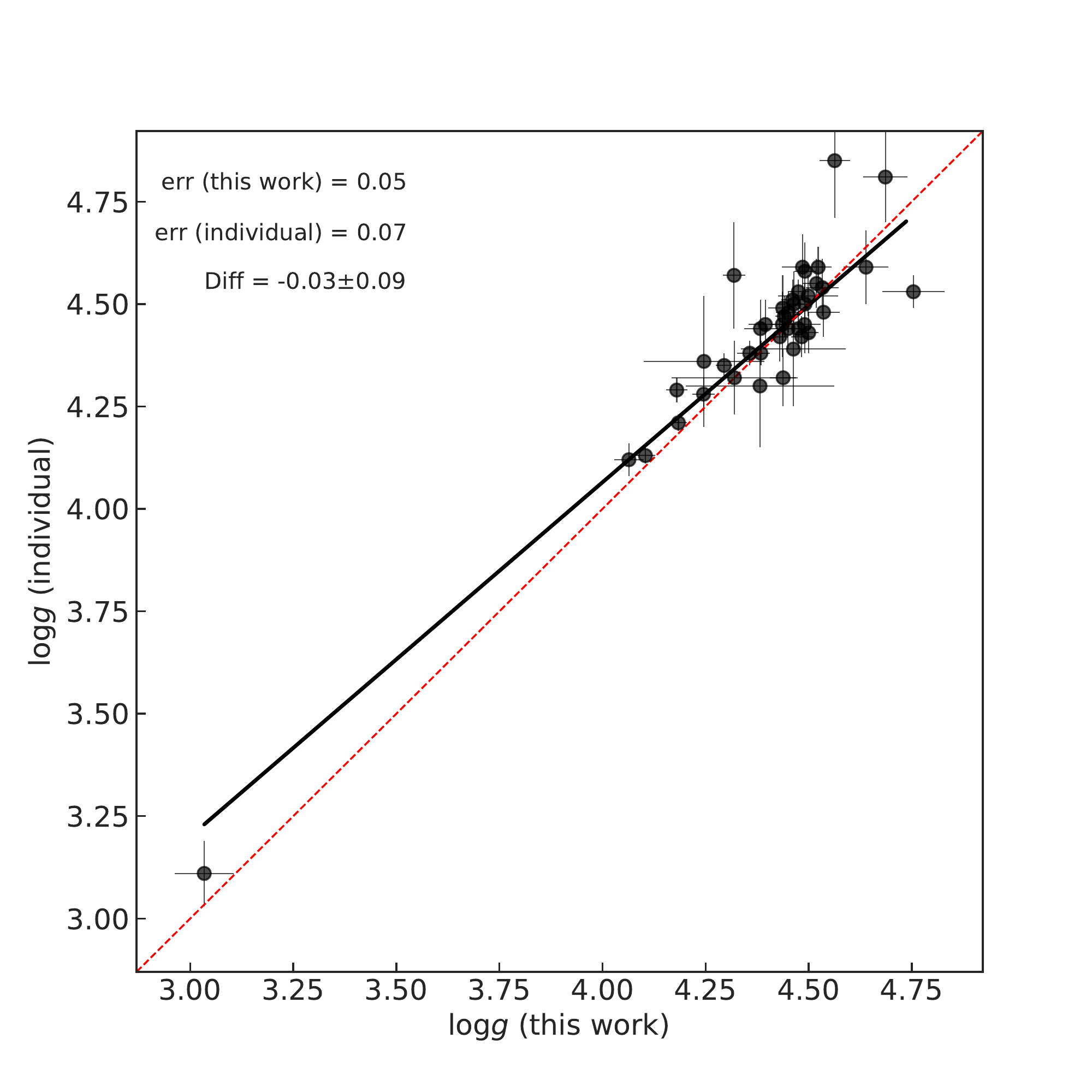} \hspace{-0.9cm}
\includegraphics[angle=0,width=0.37\linewidth]{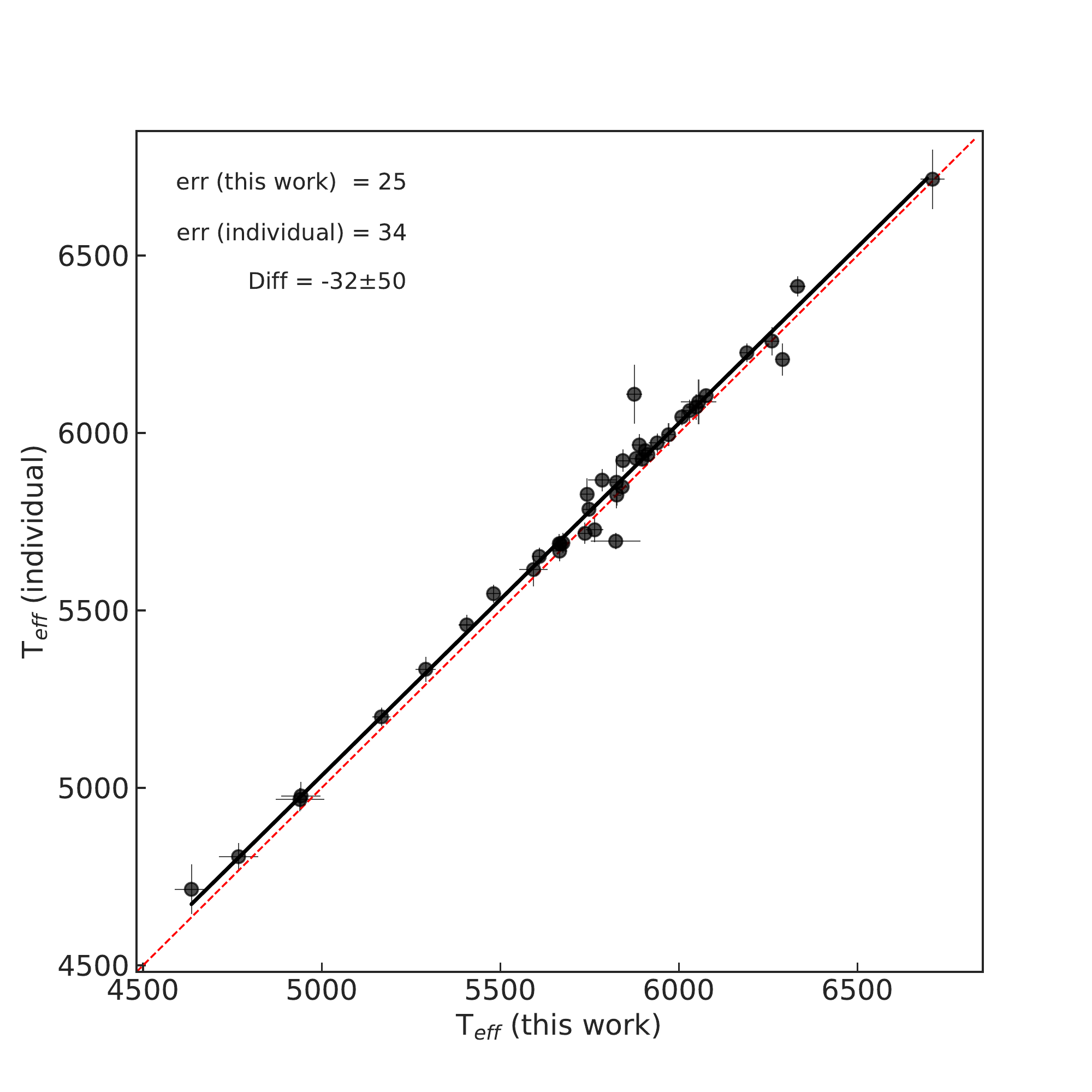} \hspace{-0.9cm}
\end{tabular}
\end{center}
\vspace{-0.5cm}
\caption{Comparison of the stellar parameters derived within the AMBRE project and here. In both works, the parameters are derived from the same individual single spectrum.}
\label{fig_slope_atmos_param}
\end{figure*}

\begin{figure*}
\begin{center}
\begin{tabular}{cc}
\includegraphics[angle=0,width=0.5\linewidth]{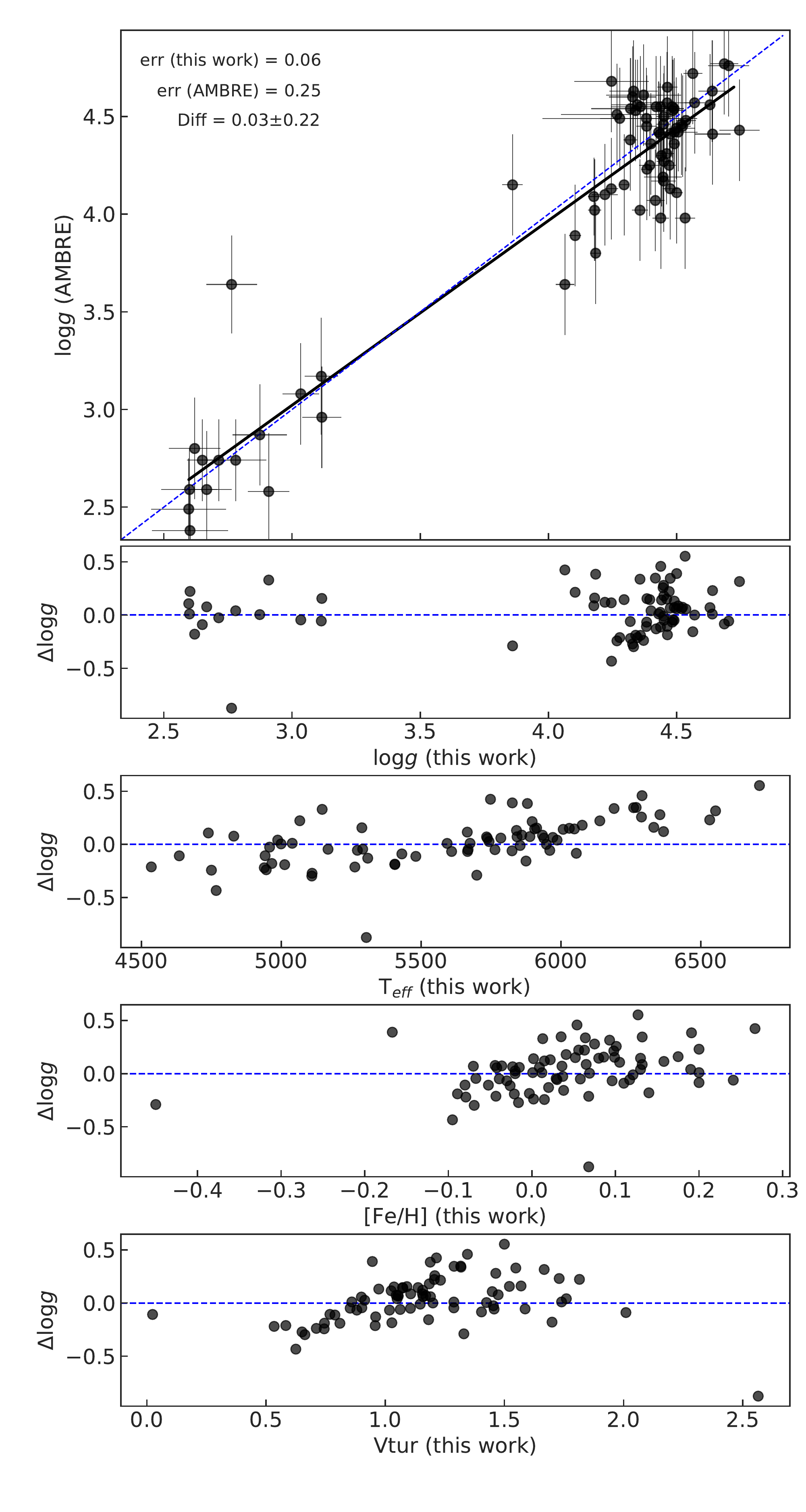}
\includegraphics[angle=0,width=0.5\linewidth]{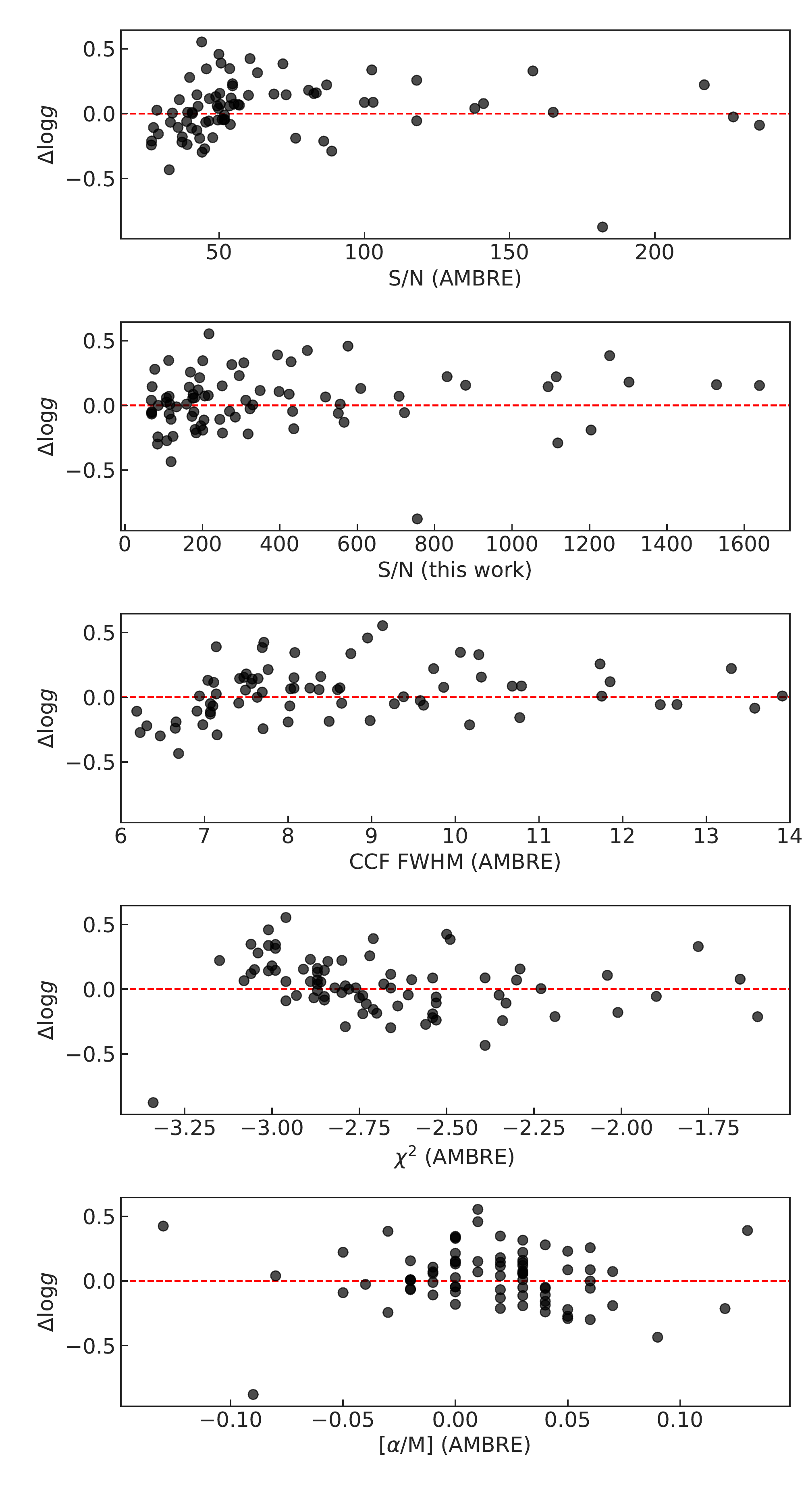}
\end{tabular}
\end{center}
\vspace{-0.6cm}
\caption{Comparison of the surface gravities derived within the AMBRE project and derived here.}
\label{fig_logg_ambre_thiswork}
\end{figure*}

\begin{figure*}
\begin{center}
\begin{tabular}{cc}
\includegraphics[angle=0,width=0.5\linewidth]{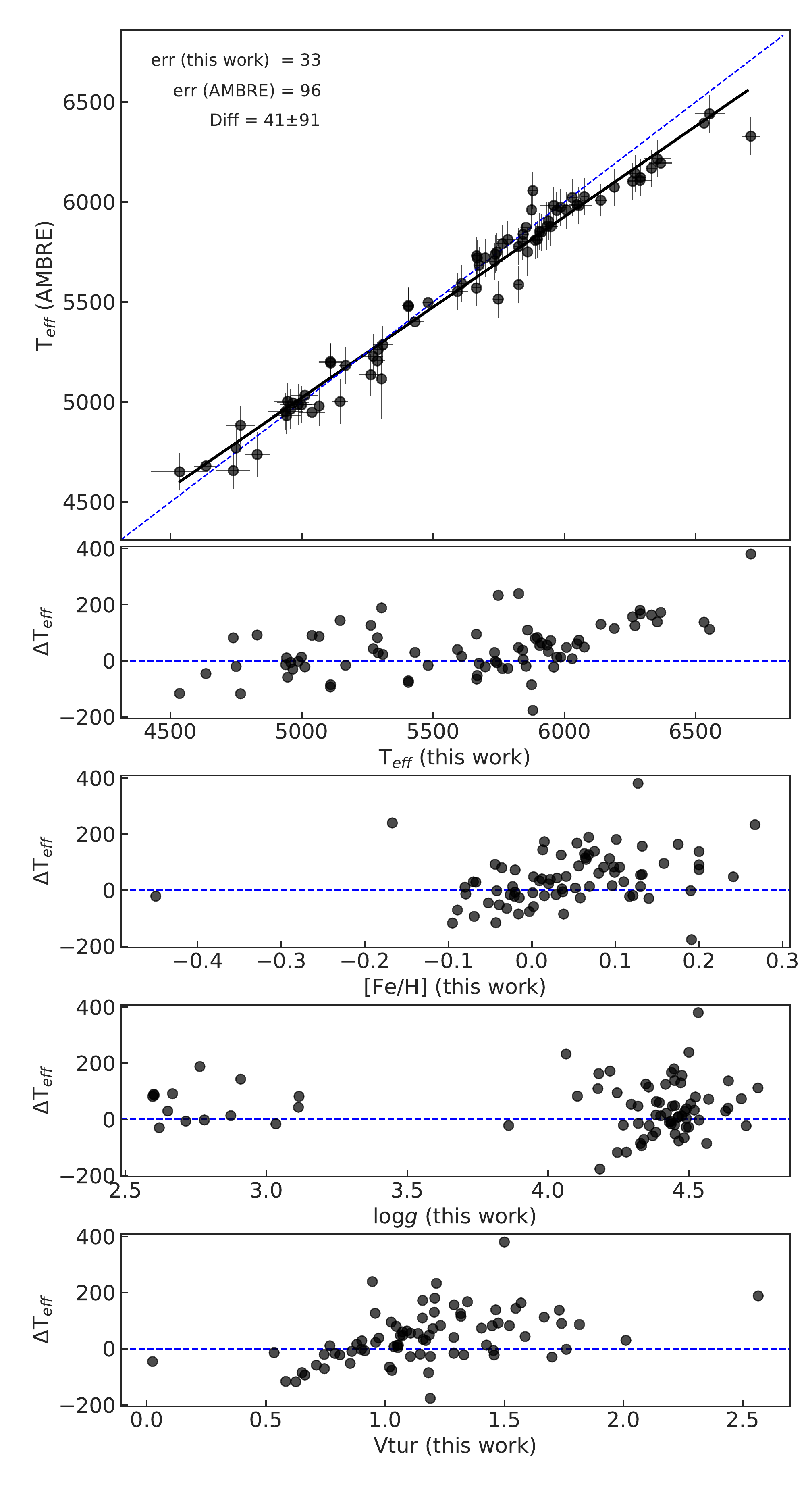}
\includegraphics[angle=0,width=0.5\linewidth]{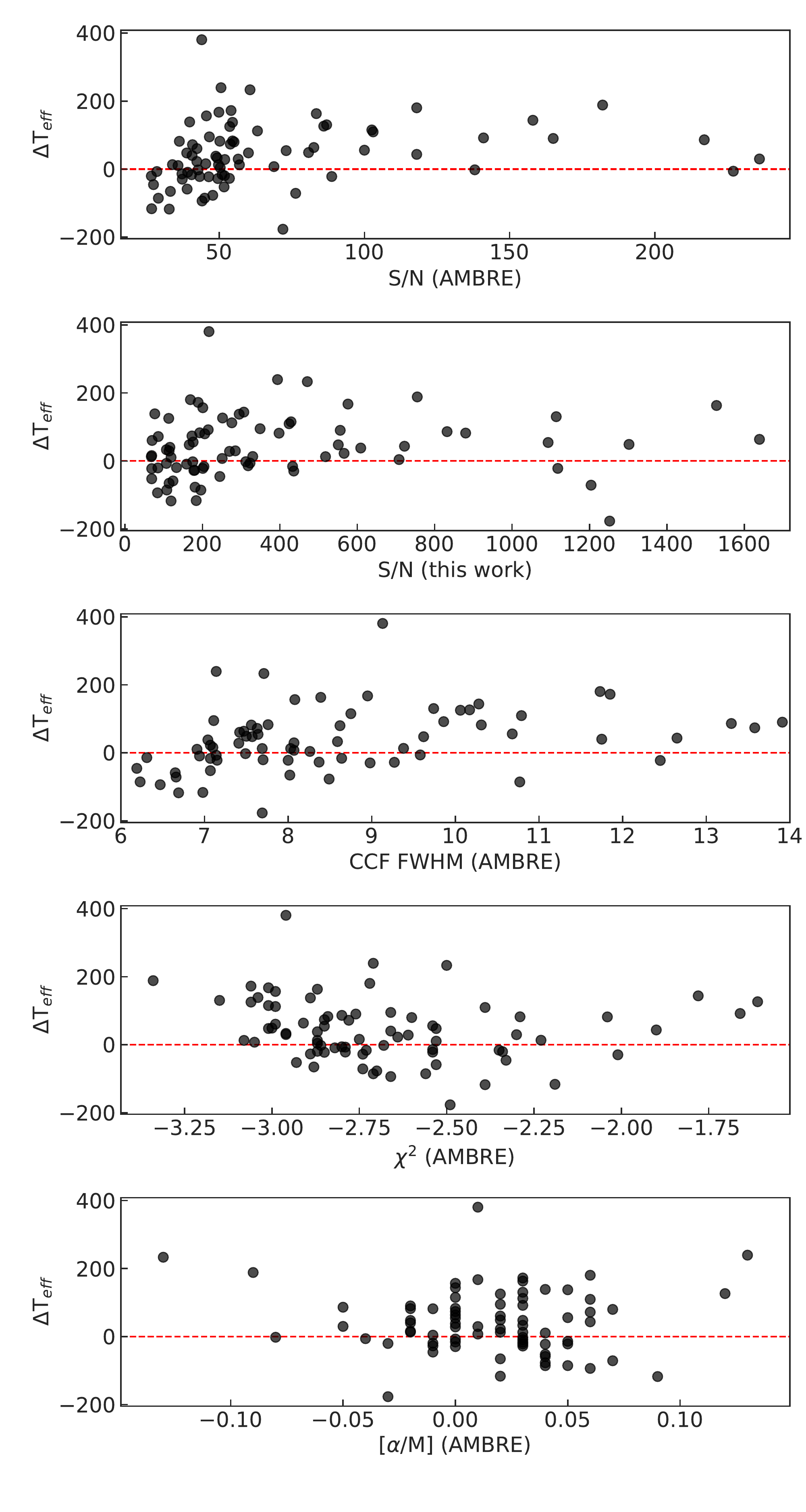}
\end{tabular}
\end{center}
\vspace{-0.6cm}
\caption{Comparison of the stellar effective temperatures derived within the AMBRE project and derived here.}
\label{fig_teff_ambre_thiswork}
\end{figure*}

\section{Abundances}

\begin{figure*}
\begin{center}
\begin{tabular}{cc}
\includegraphics[angle=0,width=0.5\linewidth]{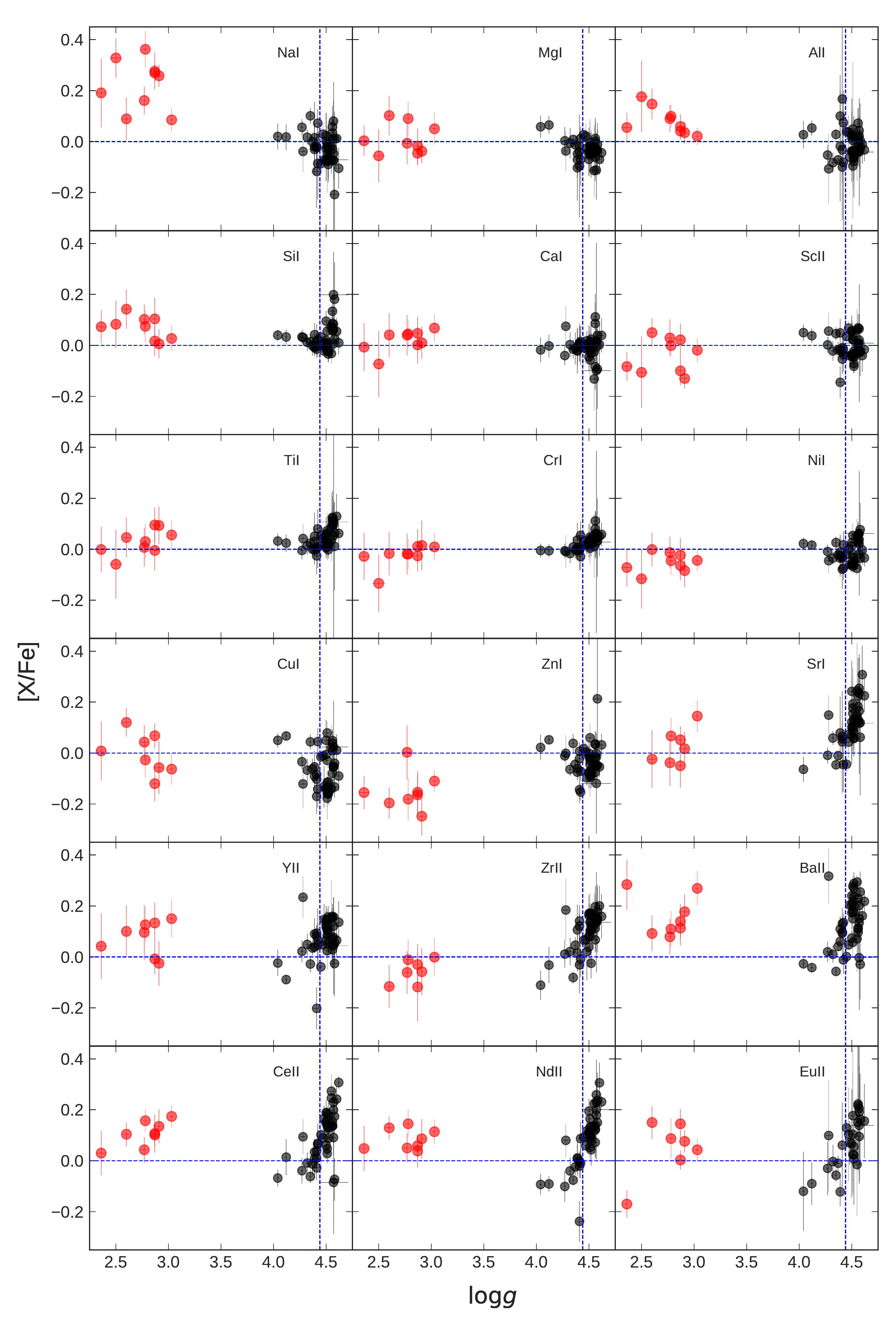}
\includegraphics[angle=0,width=0.5\linewidth]{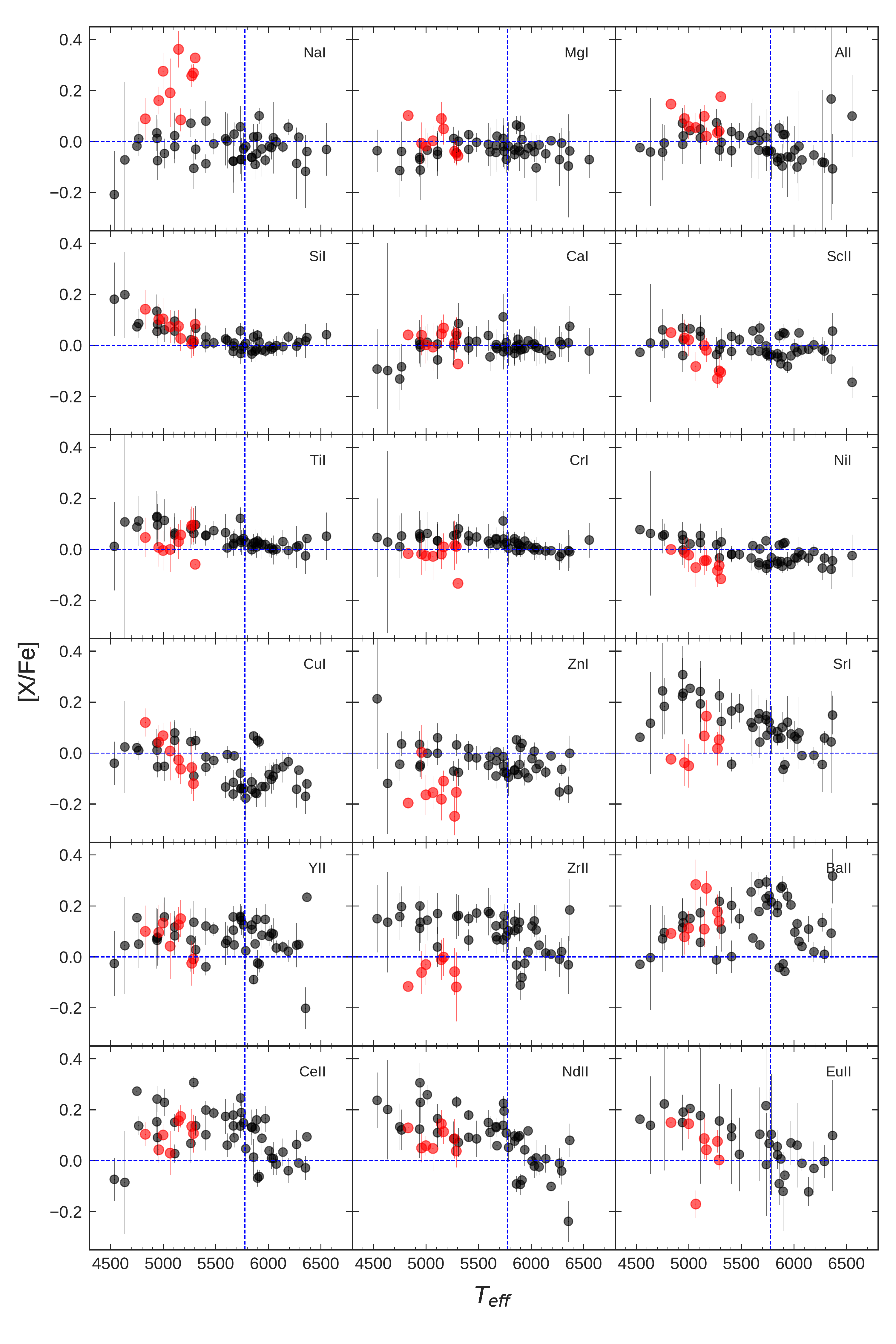}
\end{tabular}
\end{center}
\vspace{-0.6cm}
\caption{Abundance ratio [X/Fe] against trigonometric \logg \ and \teff \ for the current sample. The evolved (also massive) stars a with surface gravity lower than 3.5 dex are represented in red, and the
  stars at earlier stages of their evolution are shown in black. The blue dashed horizontal and vertical lines show the solar abundances.}
\label{fig_elfe_logg_teff}
\end{figure*}

\begin{figure*}
\begin{center}
\begin{tabular}{c}
\includegraphics[angle=0,width=0.9\linewidth]{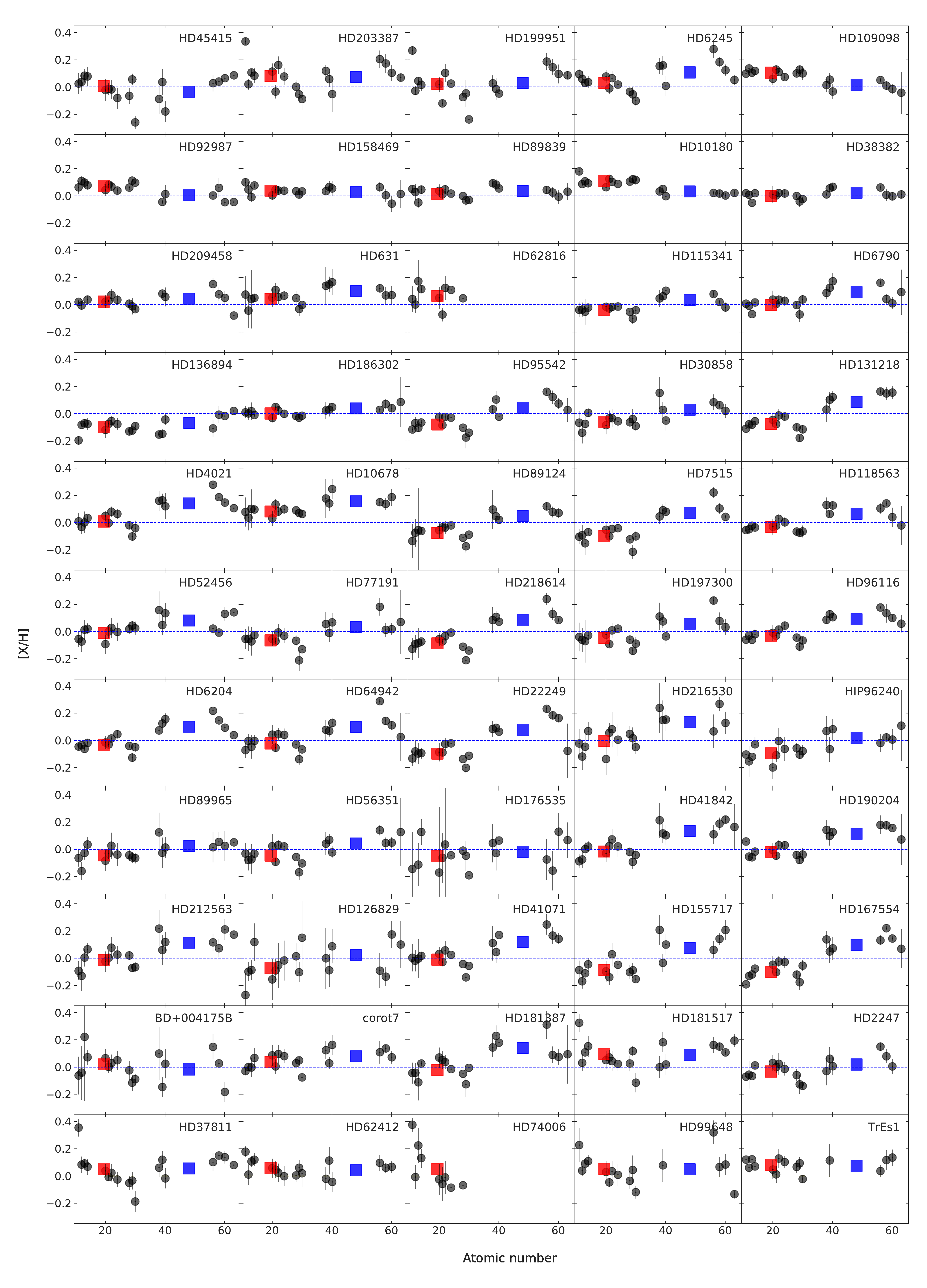}
\end{tabular}
\end{center}
\vspace{-0.6cm}
\caption{Abundances vs. atomic number for the sample stars. The average abundances of light (Z $\leqslant$ 30) and heavy (Z $>$ 30) elements are shown by red and blue squares, respectively.}
\label{fig_xh_tc_1}
\end{figure*}

\end{appendix}

\end{document}